\documentclass[preprint,showpacs,preprintnumbers,amsmath,amssymb]{revtex4}

\usepackage{graphicx}
\usepackage{epstopdf}

\usepackage{dcolumn}
\usepackage{bm}
\usepackage{color}

\usepackage{feynmf}
\usepackage{slashed}
\usepackage{enumerate}
\usepackage{caption}

\usepackage[utf8]{inputenc}

\usepackage[colorlinks=true,urlcolor=blue,anchorcolor=blue,citecolor=blue,filecolor=blue,linkcolor=blue,menucolor=blue,linktocpage=true]{hyperref} % should be commented out if the tex file will be compiled with latex in arXiv!!! (pdflatex is fine)

\begin{document}

\title{$1 \leftrightarrow 2$ Processes of a Sterile Neutrino Around Electroweak Scale in the Thermal Plasma}

\author{Xue-Min Jiang}
\affiliation{School of Physics, Sun Yat-Sen University, Guangzhou 510275, China}

\author{Yi-Lei Tang}
\thanks{tangylei@mail.sysu.edu.cn}
\affiliation{School of Physics, Sun Yat-Sen University, Guangzhou 510275, China}
%\affiliation{Quantum Universe Center, KIAS, 85 Hoegiro, Seoul 02455, Republic of Korea}

%\author{Gao-Liang Zhou}
%\thanks{zhougl@itp.ac.cn}
%\affiliation{Xi'an University of Science and Technology}

\author{Zhao-Huan Yu}
\affiliation{School of Physics, Sun Yat-Sen University, Guangzhou 510275, China}

\author{Hong-Hao Zhang}
\thanks{zhh98@mail.sysu.edu.cn}
\affiliation{School of Physics, Sun Yat-Sen University, Guangzhou 510275, China}

\date{\today}

\begin{abstract}
In this paper, we will apply the Goldstone equivalence gauge to calculate the $1 \leftrightarrow 2$ processes of a sterile neutrino in the thermal plasma below the standard model (SM) critical temperature $T_c \approx 160 \text{ GeV}$. The sterile neutrino's mass is around the electroweak scale $50 \text{ GeV} \leq m_N \leq 200 \text{ GeV}$, and the acquired thermal averaged effective width $\bar{\Gamma}_{\text{tot}}$ is continuous around the cross-over. We will also apply our results to perform a preliminary calculation of the leptogenesis.

\end{abstract}


\maketitle
\section{Introduction}

Sterile neutrinos interacting with the plasma background of the early universe 	can become a potential solution to some cosmological particle physics problems. A prominent example is the leptogenesis\cite{LeptogenesisAncester}. The CP-violation effects of the sterile neutrino interactions with the light leptons give rise to the lepton number asymmetry in the plasma, and the baryon number asymmetry accordingly appears through the sphaleron effects(for some early works, see \cite{LeptogenesisEarly1, LeptogenesisEarly2, LeptogenesisEarly3, LeptogenesisEarly4, LeptogenesisEarly5}, and see \cite{LeptogenesisReview1, LeptogenesisReview2, LeptogenesisReview3, LeptogenesisReview4} for reviews). The sterile neutrino can also become a portal to the dark matter. Being a variation of a secluded dark matter model, a ``sterile-neutrino-philic dark matter'' model\cite{MyPaper1, MyPaper2, Hantao1, Hantao2, Spain1, Signals, LateExample, MandalNew} gives a different relic density result compared with the standard weakly interacting massive particle (WIMP) models\cite{WIMPReview}. In Ref.~\cite{MyPaperBian}, we also studied a feebly interacting massive particle (FIMP)\cite{FIMPAncester} version of such kind of models. Sometimes, sterile neutrinos themselves can also become the dark matter candidate. Among all these examples, a reliable calculation of the sterile neutrino's interaction with the thermal plasma is very crucial for the precise predictions of the related physical observables compared with the experimental data.

When $m_N \gg T_c \simeq 160 \rm{ GeV}$, where $m_N$ is the sterile neutrino mass and $T_c$ is the electroweak cross-over temperature\cite{CrossOverTemp}, there are plenty of reliable discussions in the literature to calculate the sterile neutrino's production\cite{GiudiceCorrections, OverTc1, OverTc2, OverTc3, OverTc4, OverTc5, OverTc6, OverTc7, OverTc8}. Since the crucial temperature $T \sim m_N$ is well above the cross-over temperature, only the Higgs doublet and the active leptons participate the $1\leftrightarrow2$ processes. The Higgs components receive a universal thermal mass correction, which is easy to be calculated. For lighter sterile neutrinos, successful leptogenesis can also be acquired through the resonant effects\cite{Resonant1, Resonant2, Resonant3, Resonant4, ResonantEpsilon, Resonant5, Resonant6}. When $m_N \ll T_c$, at $T \sim m_N \ll T_c$, the thermal mass terms can be safely neglected since the vacuum expectation value (vev) of the Higgs boson becomes fairly close to the zero-temperature value $\sim 246 \text{ GeV}$, and the boson's behaviours are similar to those in the zero-temperature situation\cite{SterileNeutrinoDecay1}.

In the literature, there seems to be a gap when $m_N \sim T_c$. In this range the calculation is plagued by the intricate thermal corrections to the gauge and Higgs sectors. In Ref.~\cite{ThomasPRL}, the authors estimated the U(1)$_Y\times$SU(2)$_L$ gauge boson contributions by replacing them with the Goldstone degrees of freedom artificially assigned with the similar mass of the Higgs boson. We also applied this method in the corresponding calculations of our papers \cite{MyPaper2, MyPaperBian}. Such an ansatz might be inspired by the famous ``Goldstone equivalence theorem'' in the zero temperature, which requires more investigations in the thermal plasma case. A safe procedure is to return to the original form of the finite temperature propagators to integrate all the branch cuts and poles whatever appear, as described in Ref.~\cite{BelowTcLaine1, BelowTcLaine2, BelowTcLaine3, BelowTcLaineNLO}. However, it is formidable for one to follow the procedures there, and the relationship between the Goldstone and gauge boson becomes more obscure. Another fact is that the invariant squared mass of the sterile neutrino, which is denoted by $\mathcal{K}^2$ in Ref.~\cite{BelowTcLaine1, BelowTcLaine2, BelowTcLaine3, BelowTcLaineNLO}, had been neglected around $T_c$ there, so their method is not suitable to our interested $\mathcal{K}^2 = m_N^2 \sim T_c^2$ range.

In Ref.~\cite{VectorBosonMyPaper} we proposed a method to decompose the massive gauge boson propagators in the thermal plasma. Poles indicating the ``transverse'' and ``longitudinal'' degrees of freedom arise as usual, and a branch cut which extremely resembles two massless poles was identified as the Goldstone boson's fragment. When $T>T_c$, such a branch cut fragments into two actual poles corresponding to the Goldstone boson particles, and when $T=0$, this branch cut completely disappears. In the finite temperature, the longitudinal polarization is also some intermediate state between the so-called ``plasmon'' and the Goldstone equivalent state. We made an analogy that the longitudinal polarization will ``spew out'' a fraction of the Goldstone boson in the finite temperature environment. This helps us include all the contributions from the transverse, longitudinal, Higgs and Goldstone degrees of freedom correctly, and help us clarify the relationship between the Goldstone and the gauge bosons in the plasma.

In this paper, with the method we have developed in Ref.~\cite{VectorBosonMyPaper}, we will calculate the sterile neutrino $1\leftrightarrow2$ processes near the electroweak cross-over temperature $m_N \sim T \sim T_c$. We will also roughly discuss the leptogenesis induced by these processes. A complete calculation of the sterile neutrino's interaction in the early universe should also include the more complicated $2 \leftrightarrow 2$ scattering processes. In many cases when $T \gg m_N$, and the $l$-$H$-$N$ Yukawa couplings $y_N \gtrsim 10^{-8}$ which are sufficiently large, thermal equilibrium of the sterile neutrino does not require a detailed calculation. When the temperature drops down to the $T \sim m_N$ scale, the out-of-equilibrium effects start to arise, and these $2 \leftrightarrow 2$ processes are usually suppressed rapidly due to an additional number density factor compared with the $1 \leftrightarrow 2$ processes. With these considerations, we leave the $2 \leftrightarrow 2$ processes to our future study and do not consider their contributions on this stage. We also do not consider the contributions resumming the interchange/emission of the soft bosons\cite{LPM1, LPM2, LPM3} (sometimes called the LPM resummation) in this paper for brevity and simplicity.

We enumerate the channels and list the basic formulas in Sec.~\ref{Enumeration}. Details on phase space and thermal integrals are presented in Sec.~\ref{Integration}. Numerical results and a preliminary calculation of leptogenesis are displayed in Sec.~\ref{NumericalResults}. We summarize this paper in Sec.~\ref{Summery}.

\section{Basic Concepts and Channel Enumeration}	\label{Enumeration}

The Lagrangian of sterile neutrino is the standard one
\begin{eqnarray}
\mathcal{L} \supset \mathcal{L}_{\rm{SM}} + \mathcal{L}_{\text{N kin}} + \mathcal{L}_{\text{N mass}} - \sqrt{2} y_{N i j} H \bar{l}_i N_j + \text{h.c.},
\end{eqnarray}
where $H$ is the Higgs doublet, $L_i$, $i=1,2,3$ are the lepton doublets of three generations, $N_j$ are the sterile neutrinos. $N_j$ can be either Majorana or (pseudo-)Dirac spinors, and the corresponding kinematical and mass terms $\mathcal{L}_{\text{N kin}} + \mathcal{L}_{\text{N mass}}$ differ by a factor of $\frac{1}{2}$. For simplicity here we only study the one Dirac sterile neutrino case. The interaction only involves one massless lepton. A general situation can be inferred from our results by simply multiplying some factors. Therefore, the Lagrangian we are relying on is given by
\begin{eqnarray}
\mathcal{L} \supset \mathcal{L}_{\rm{SM}} + i \bar{N} \partial\!\!\! / N - m_N \bar{N} N - \sqrt{2} y_{N} H \bar{l} N + \text{h.c.},
\end{eqnarray}
where $m_N$ is the mass of the sterile neutrino.

Above the standard model (SM) critical temperature of the cross-over $T > T_c \approx 160 \text{ GeV}$, the $1\leftrightarrow2$ processes of the sterile neutrino have nothing to do with the W/Z boson. Only the Higgs doublets including the Goldstone components participate the couplings. The whole process is quite standard: the thermal effects correct the effective Higgs mass term
\begin{eqnarray}
\delta m_{H,\text{ thermal}}^2 = (g_1^2+3 g_2^2 + 4 y_t^2 + 8 \lambda) \frac{T^2}{16},
\end{eqnarray}
where $g_1$, $g_2$ are the electroweak gauge coupling constants, $y_t$ is the top Yukawa coupling constant, and the $\lambda$ is the 4-Higgs coupling constant. Leptons also receive the thermal mass corrections. In the thermal plasma, each pole in the leptonic propagators are split into two objects, so called a ``particle'' and a ``hole''. In the Ref.~\cite{GiudiceCorrections}, both these two objects are combined into one single particle with the universal thermal mass correction to estimate the phase space. In this paper, we abandon this approximation, and earnestly sum over each contributions from these two degrees of freedom.

Below the critical temperature $T<T_c$, the vacuum expectation value (vev) is estimated to be $v(T)=v_0 \sqrt{1-\frac{T^2}{T_c^2}}$, where $v_0 = 246 \text{ GeV}$. This opens the sterile neutrino's oscillation into a highly off-shell active neutrino, and then it decays into a W/Z gauge boson plus a charged lepton/active neutrino. An on-shell W/Z boson can also decay into a pair of leptons, and the active neutrino product can also oscillate into a sterile neutrino through the vev.

The dispersion relations (or the ``on-shell'' equation) of the W/Z bosons below the critical temperature are complicated. Together with the dispersion relations of the leptons, and the conservation of energy and momentum equations, we have four equations to solve the phase space. Three of them are transcendental equations. Later we are going to describe the details to solve them.

In this paper, we rely on the Goldstone equivalent gauge\cite{GoldstoneEquivalenceGauge} to calculate the sterile neutrino's productions in the thermal plasma\cite{VectorBosonMyPaper} below the critical temperature $T_c$. Within this framework, each Goldstone degree of freedom is attributed into two parts: one is hidden inside the extended polarization vector of a longitudinal vector boson, another behaves like a massless particle during the calculations, and is regarded independently as a Goldstone boson's fraction. We enumerate and include all of the gauge polarizations and the Goldstone boson fraction's contributions.  In the appendix, we will also show the equivalence between this gauge and the usually familiar $R_{\xi}$ gauge.

In the following subsections we will describe the details for each channel. Before starting them, we also note that we ignore some of the sub-dominant tachyonic branch cuts in the bosonic propagators, as illustrated in our Ref.~\cite{VectorBosonMyPaper}, and as in Ref.~\cite{GiudiceCorrections}, the sub-dominant branch cuts in the leptonic propagators are also neglected.

\subsection{W channels}	\label{WChannelSection}

\begin{figure}
\includegraphics[width=2in]{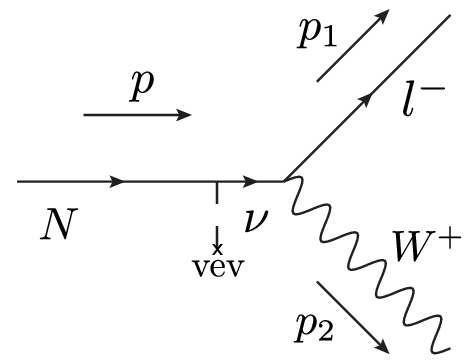}
\includegraphics[width=2in]{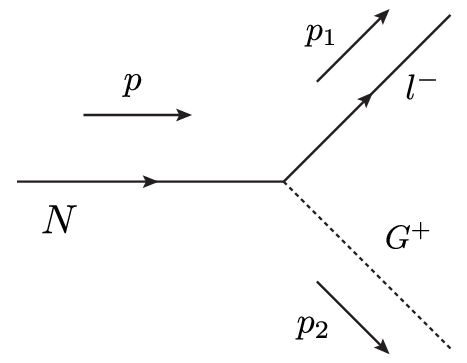}
\caption{$N \rightarrow W^+ l^-$ $1\leftrightarrow2$ channel. Since we have applied the Goldstone equivalence gauge, the Goldstone contribution is explicitly contained in the polarization vector, so we also need to calculate the Goldstone part of the diagrams.} \label{NWl}
\end{figure}

The Feynmann diagram of a sterile neutrino $N$ decaying into a $W^+$ boson and a charged lepton $l^-$ is illustrated in Fig.~\ref{NWl}. Since we are discussing a Dirac $N$, it is possible to inverse the arrows there to reformulate it into a $\overline{N}$ decay diagram. We neglect the anti-sterile neutrino's decay in our paper since the results are completely symmetric by neglecting the CP effects. The momentum flows are also defined in Fig.~\ref{NWl} and are defined relative to the plasma background reference, i.e., the plasma's four-vector velocity 
\begin{eqnarray}
u^{\mu} = (1,0,0,0).
\end{eqnarray}
When, e.g., $p_1^0<0$, the same diagram can also be interpreted as a charged lepton's fusion with the sterile neutrino to generate a $W^+$ boson, which is the dual process of a $W^+$ decaying into a $N$, $l^+$ pair. This is the ``inverse-decay'' process of a $W^+$ boson, and we denote it with ``ID'' for abbreviation later. The thermal equilibrium condition guarantees the equality of the results from both the aspects of ``decay'' and ``inverse-decay'' processes of a $W$ boson. Therefore, Fig.~\ref{NWl} can summarize all the possible $1\leftrightarrow2$ processes of a (anti-)sterile neutrino. 

The dispersion relation of a W boson is given by
\begin{eqnarray}
F_{W,(L,T)} (p_2) = p_2^2-[m_W(T)]^2 -\Pi_{L,T}^W(p_2)=0,		\label{WDispersion}
\end{eqnarray}
for transverse and longitudinal polarizations respectively, where
\begin{eqnarray}
\Pi_L^W (p_2) &=& -\frac{2 m_{E2}^2 p_2^2}{\vec{p}_2^2} \left( 1- \frac{p_2^0}{|\vec{p}_2|} Q_0(\frac{p_2^0}{|\vec{p}_2|}) \right), \nonumber \\
\Pi_T^W(p_2) &=& \frac{1}{2}(2 m_{E2}^2 - \Pi_L^W(p_2) ), \label{PiW}
\end{eqnarray}
and
\begin{eqnarray}
Q_0(x) = \frac{1}{2} \ln \frac{x+1}{x-1}.
\end{eqnarray}
The vev dependent W boson mass is given by
\begin{eqnarray}
m_W (T) = \frac{g_2 v(T)}{2},
\end{eqnarray}
where $g_2$ is the weak coupling constant, and the Debye thermal mass $m_{E2}$ takes the form
\begin{eqnarray}
m_{E2}^2 = \frac{11}{6} g_2^2 T^2.
\end{eqnarray}

Ignoring the lepton's vev dependent mass, since it is much smaller than the thermal mass term, the thermal corrected dispersion relation of the active lepton is given by(See page 140 in Ref.~\cite{ThermalFieldBook})
\begin{eqnarray}
F_{l} (p_1) =\left[ \Delta_{+} (p_1) \Delta_{-}(p_2) \right]^{-1} = 0,	\label{LDispersion}
\end{eqnarray}
where
\begin{eqnarray}
\Delta_{\pm}(p_1) = \left(p_1^0 \mp |\vec{p}_1| - \frac{m_f^2}{2 |\vec{p}_1|} \left[ \left( 1 \mp \frac{p_1^0}{|\vec{p}_1|} \right) \ln \frac{p_1^0+|\vec{p}_1|}{p_1^0-|\vec{p}_1|} \pm 2 \right] \right)^{-1}.
\end{eqnarray}
Here
\begin{eqnarray}
m_f^2 = \frac{g_1^2 + 3 g_2^2}{32} T^2.	\label{FermionCorrectionMass}
\end{eqnarray}
Generally there are four solutions to the ($\ref{LDispersion}$). When $p_1^2 > m_f^2$, this means a ``particle''  for $p_1^0>0$, and an ``anti-particle'' for $p_1^0<0$. When $p_1^2 < m_f^2$, this indicates a ``hole'' for $p_1^0>0$, and an ``anti-hole'' for $p_1^0<0$.

The energy and momentum conservation laws are given by
\begin{eqnarray}
p^0 &=& p_1^0 + p_2^0, 	\label{EConserve} \\
\vec{p}_2^2 &=& \vec{p}^2 + \vec{p}_1^2 - 2 |\vec{p}||\vec{p}_1| \cos \theta_p,	\label{pConserve}
\end{eqnarray}
where $\theta_p$ is the angle between $\vec{p}$ and $\vec{p_1}$. The subscript ``$p$'' denotes the ``plasma'', which means that this is the angle measured in the plasma rest frame. Given the sterile neutrino's energy and momentum $p^0$, $\vec{p}$, fixing the $\theta_p$, there are four unknown parameters $p_1^0$, $p_2^0$, $|\vec{p}_1|$, $|\vec{p}_2|$ in just four equations (\ref{WDispersion}, \ref{LDispersion}, \ref{EConserve}, \ref{pConserve}). Solving these equations might give a set of solutions. If $p_1^0$ or $p_2^0$ is smaller than zero, it means that a lepton or a W boson becomes an initial state particle. We need to find all of the solutions to sum over all their contributions to the ``interaction rate'' $\gamma_N$.

With the acquired $p_1$ and $p_2$, we can then calculate the amplitude. In the Goldstone equivalence gauge, the ``polarization vector'' of a gauge boson is extended to a five-component vector $\epsilon_{\pm,L\text{in}}^{Wn}(p_2) = \epsilon_{\pm,L\text{out}}^{Wn*}(p_2)$, $n=\mu,4$ to include the Goldstone component ($n=4$ denotes the Goldstone component). When contracting the indices, the metric tensor $[g^{\mu \nu}] = \text{diag}[1,-1,-1,-1]$ is also extended to $[g^{m n}] = \text{diag}[1,-1,-1,-1,-1]$. The transverse polarization is the same as in the $R_{\xi}$ gauge with $\epsilon_{\pm}^{W4}(p_2)=\epsilon_{\pm}^{W0}(p_2)=0$, and $\epsilon_{\pm}^{Wi}(p_2) p_{2i} = 0$. The longitudinal polarization $\epsilon_{L\text{in}}^{Wn*} (p_2)=\epsilon_{L\text{out}}^{Wn} (p_2)$ is given by
\begin{eqnarray}
\epsilon_{L\text{out}}^W(p_2) = \left( 
\begin{array}{c}
-\frac{\sqrt{p_2^2}}{n_2 \cdot p_2} n^{\mu}_2 \\
-i \frac{m_{W}(T)}{\sqrt{p_2^2}}
\end{array}
\right), \label{LongitudinalPolarization}
\end{eqnarray}
where $n_{2}^{\mu} = (1, -\frac{\vec{p}_2}{|\vec{p}_2|})$ for the convention of $(k^{\mu}) = (k^0, \vec{k})$ for any four-dimensional momentum $k$.

For the lepton spinors, we need to define
\begin{eqnarray}
\tilde{p}_1 = p_1^0 (1, \pm \frac{\vec{p}_1}{|\vec{p}_1|}),
\end{eqnarray}
where for a ``particle'', i.e., $p_1^2 > m_f^2$, the ``+'' sign is adopted, and for a ``hole'', i.e., $p_1^2 < m_f^2$, the ``-'' sign is adopted. When $p_1^0>0$, a lepton (either a ``particle'' or a ``hole'') is created and a $\bar{u}^s(\tilde{p}_1)$ appears in the amplitude. When $p_1^0<0$, an anti-lepton (either an anti-``particle'' or an anti-``hole'') is destroyed and a $\bar{v}^s(-\tilde{p}_1)$ appears in the amplitude respectively.

The amplitude of the gauge component, as denoted in the left panel of Fig.~\ref{NWl}, then becomes
\begin{eqnarray}
i \mathcal{M}^{\mu}_{W} = -y_N v(T) \frac{g_2}{\sqrt{2}} \bar{u}^s (\tilde{p}_1) \left[ \gamma^{\mu}+\Gamma^{\mu}(p, p_1) \right] P_L \frac{i}{p\!\!\!/_{lT}} u^r (p), \label{MWV}
\end{eqnarray}
when $p_1^0 > 0$ for the decay channel. {$\Gamma^{\mu}(p, p_1)$ is the HTL correction on gauge vertex  introduced for a gauge invariant result. Its definition is  given in  (\ref{GaugeVertexCorrection}), followed by the detailed evaluation processes there in the appendix.} If $p_1^0<0$, we only need to change the $\bar{u}^s (\tilde{p}_1)$ into $\bar{v}^s (-\tilde{p}_1)$ for the $W$-boson's inverse-decay channel. The Goldstone component of the amplitude as denoted in the right panel of Fig.~\ref{NWl}, is written to be
\begin{eqnarray}
i \mathcal{M}^4_W = -\sqrt{2}  y_N \bar{u}^s (\tilde{p}_1) P_R u^r (p). \label{MWGS}
\end{eqnarray}
Again when $p_1^0<0$, $\bar{u}(\tilde{p}_1)$ needs to be replaced with $\bar{v}(-\tilde{p}_1)$. In the above equations, $P_{L,R} = \frac{1 \mp \gamma^5}{2}$, and the definition of $p_{lT}$ is
\begin{eqnarray}
p_{lT} = (p_{lT}^0, \vec{p}_{lT}) = ( (1-\frac{m_f^2 L}{p^0})p^0, (1+\frac{m_f^2 (1-p^0 L)}{\vec{p}^2}) \vec{p}),
\end{eqnarray}
where
\begin{eqnarray}
L = \frac{1}{2 |\vec{p}|	} \ln \frac{p^0+|\vec{p}|}{p^0-|\vec{p}|}.
\end{eqnarray}
The complete amplitude should take the form
\begin{eqnarray}
\epsilon_{(t) n}^W(p_2) (i \mathcal{M}_W^{n}),
\end{eqnarray}
where $n=0,1,2,3,4$, $t=\pm, L\text{out}$. The squared amplitude should also take the statistic factor and the ``renormalization constant''. The complete result is
\begin{eqnarray}
A_{W,t} = \sum_{r,s=1,2} \mathcal{M}_W^{n} \mathcal{M}_W^{* m} \epsilon_{t n}^W \epsilon^{W*}_{t m} f_F(\frac{p_1^0}{T} ) f_B(\frac{p_2^0}{T}) Z_l(p_1) Z_{W t}(p_2), \label{WTotalSquaredAmplitude}
\end{eqnarray}
where $t=\pm, L$ indices are not summed by the Einstein's sum rule, and
\begin{eqnarray}
f_F(x) &=& \frac{e^{x}}{e^{x}+1}, \\
f_B(x) &=& \left| \frac{e^x}{e^x-1} \right|,
\end{eqnarray}
and the ``renormalization factors'' are
\begin{eqnarray}
Z_{W (T, L)}(p_2) &=& \frac{2 p_2^0}{2 p_2^0 - \frac{\partial \Pi_{T,L}^{W}(p_2)}{\partial p_2^0}}, \\
Z_{l}(p_1) &=& \frac{ (p_1^0)^2-\vec{p}_1^2}{2 m_f^2}. \label{LeptonRenormalizationFactor}
\end{eqnarray}

\subsection{$Z$/$\gamma$ channels}

Since $W$ and $B$ bosons receive the different thermal corrections, it disturbs the mixing angle for the ``on-shell'' $Z$/$\gamma$ bosons. The mixing angles of the on-shell $Z$/$\gamma$ bosons depend on their energy and momentum, so it is difficult to identify which is the $Z$ or $\gamma$ degree of freedom.

\begin{figure}
\includegraphics[width=2in]{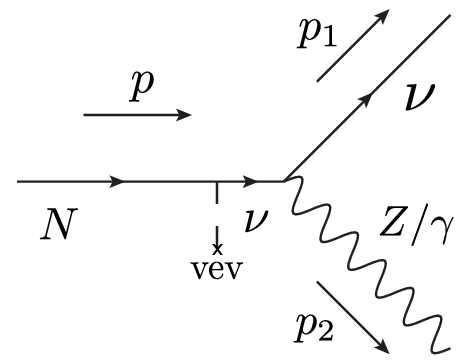}
\includegraphics[width=2in]{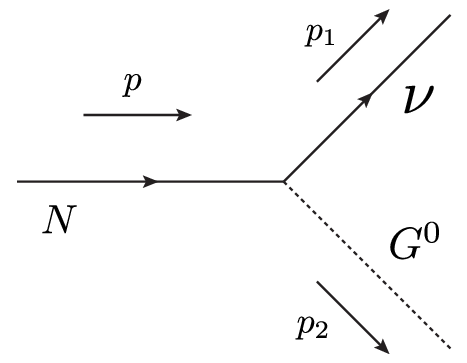}
\caption{$N \rightarrow (Z\text{/}\gamma) \nu$ $1\leftrightarrow2$ channel.} \label{NZv}
\end{figure}

The vev dependent mass matrix for the $B$/$W^3$ field, or $Z$/$\gamma$ particle is as usual
\begin{eqnarray}
m_{Z\text{/}\gamma}^2 (T) = \frac{(v(T))^2}{4} \left( \begin{array}{cc}
g_1^2 & -g_1 g_2 \\
-g_1 g_2 & g_2^2
\end{array} \right).
\end{eqnarray}
Thermal effects correct the $B$ and $W^3$ mass terms respectively, and therefore the thermal mass matrix is given by
\begin{eqnarray}
\Pi_{T,L}^{Z\text{/}\gamma} (p_2)= \left( \begin{array}{cc}
\Pi_{T,L}^{B} (p_2) & 0 \\
0 & \Pi_{T,L}^{W} (p_2)
\end{array} \right),
\end{eqnarray}
where $\Pi_{T,L}^{W}(p_2)$ had already been given by (\ref{PiW}). $\Pi_{T,L}^{B}$ changes the $m_{E2}$ in (\ref{PiW}) into $m_{E1}$,
\begin{eqnarray}
m_{E1}^2 = \frac{11}{6} g_1^2 T^2.
\end{eqnarray}
The dispersion rate of this mixed $Z$/$\gamma$ is given by the ``secular equation''
\begin{eqnarray}
F_{Z\text{/}\gamma, (T,L)}(p_2) = \det ( p_2^2 I_{2 \times 2} - m_{Z\text{/}\gamma}^2(T) - \Pi_{T,L}^{Z\text{/}\gamma} (p_2)) = 0,	\label{DispersionZGamma}
\end{eqnarray}
for a transverse/longitudinal $Z$/$\gamma$ vector boson. $I_{2 \times 2}$ is the $2 \times 2$ identity matrix. For a given $p_2$ as a solution of (\ref{DispersionZGamma}), matrix  $p_2^2 I_{2 \times 2} - m_{Z\text{/}\gamma}^2(T) - \Pi_{T,L}^{Z\text{/}\gamma} (p_2)$ has a zero eigenvalue, and the corresponding eigenvector is denoted by $x=\left( \begin{array}{c} x_1 \\ x_2 \end{array}\right) $, where $x_1^2+x_2^2=1$. In the zero temperature case, $x_1^Z = -\sin \theta_W$, $x_2^Z=\cos \theta_W$ for the $Z$ boson, and $x_1^{\gamma}=\cos \theta_W$, $x_2^{\gamma}=\sin \theta_W$ for the photon, where $\theta_W$ is the Weinberg angle. Since the neutrino does not interact with a pure photon, we can calculate the inner product $x \cdot x^Z = -x_1 \sin \theta_W + x_2 \cos \theta_W$ to extract the $Z$ part of the ``on-shell' mixed boson to calculate its interactions with the leptons. The dispersion relation of a lepton and the energy-momentum conservation law is exactly the same with (\ref{LDispersion}, \ref{EConserve}, \ref{pConserve}) in Sec.~\ref{WChannelSection}. Solve these equations with (\ref{DispersionZGamma}), we then acquire all the ``on-shell'' $p_1$ and $p_2$.

The transverse polarization vectors of a $Z$/$\gamma$ boson $\epsilon_{\pm}^{Z\text{/}\gamma n}$ is the same as the W-boson $\epsilon_{\pm}^{Wn}$ to satisfy $p_{2 \mu} \epsilon_{\pm}^{Z \text{/} \gamma \nu} = 0$, $\epsilon_{\pm}^{Z \text{/} \gamma 4} = 0$ and $p_{2 i} \epsilon_{\pm}^{Z \text{/} \gamma i} = 0$. The longitudinal polarization vector is given by
\begin{eqnarray}
\epsilon_{L\text{in}}^{Z\text{/}\gamma *}(p_2) = \epsilon_{L\text{out}}^{Z\text{/}\gamma}(p_2) = \left( 
\begin{array}{c}
-\frac{\sqrt{p_2^2}}{n_2 \cdot p_2} n^{\mu}_2 \\
-i \frac{m_{Z}(T)}{\sqrt{p_2^2}} (-x_1 \sin \theta_W + x_2 \cos \theta_W)
\end{array}
\right). \label{LongitudinalPolarizationZGamma}
\end{eqnarray}
Compared with the (\ref{LongitudinalPolarization}), the extra $(-x_1 \sin \theta_W + x_2 \cos \theta_W)$ factor in the Goldstone component indicates that only the $Z$-component of the vector boson had ``eaten'' some Goldstone boson. The photon part of this vector boson had not devoured any Goldstone boson's fraction.

Then we are ready to write the amplitudes.
\begin{eqnarray}
i \mathcal{M}_{Z\text{/}\gamma}^{\mu} &=& -y_N v(T) \frac{g_2}{2 \cos \theta_W} \bar{u}^s (\tilde{p}_1) \left[\gamma^{\mu} + \Gamma^{\mu}(p, p_1)\right] P_L \frac{i}{p\!\!\!/_{lT}} u^r (p)(-x_1 \sin \theta_W + x_2 \cos \theta_W), \label{MZV}\\
i \mathcal{M}^4_{Z\text{/}\gamma} &=& -	 y_N \bar{u}^s (\tilde{p}_1) P_R u^r (p).	\label{MZGS}
\end{eqnarray}

The total result of the squared amplitude is
\begin{eqnarray}
A_{Z\text{/}\gamma, t} = \sum_{r,s=1,2} \mathcal{M}_{Z\text{/}\gamma}^{n} \mathcal{M}_{Z\text{/}\gamma}^{* m} \epsilon_{t n}^{Z\text{/}\gamma} \epsilon^{{Z\text{/}\gamma}*}_{t m} f_F(\frac{p_1^0}{T} ) f_B(\frac{p_2^0}{T}) Z_l(p_1) Z_{{Z\text{/}\gamma} t}(p_2),
\end{eqnarray}
where the ``renormalization constant'' $Z_{Z\text{/}\gamma (T/L\text{out})} (p_2)$ is calculated to be
\begin{eqnarray}
Z_{{Z\text{/}\gamma} (T, L)}(p_2) &=& \frac{2 p_2^0}{2 p_2^0 -  \Pi_{(T,L), p_2^0}^{Z\text{/}\gamma,\text{ on shell}}(p_2)},
\end{eqnarray}
and $\Pi_{(T,L),p_2^0}^{Z\text{/}\gamma,\text{ on shell}}(p_2) = x^T \frac{\partial \Pi_{T,L}^{Z\text{/}\gamma} (p_2)}{\partial p_2^0} x$ so that
\begin{eqnarray}
\Pi_{T,L}^{Z\text{/}\gamma,\text{ on shell}}(p_2) = x_1^2 \frac{\partial \Pi_{T,L}^B(p_2)}{\partial p_2^0} + x_2^2 \frac{\partial \Pi_{T,L}^W(p_2)}{\partial p_2^0}.
\end{eqnarray}

\subsection{Goldstone channels}

Besides the Goldstone components in the $Z_L$ and $W_L$ polarization vectors, the Goldstone boson's fragments also contribute to the $1\leftrightarrow2$ rate. Rigorously speaking these remains are no longer a ``particle'' since they are ``branch cuts'' rather than ``poles''. However, since the imaginary parts peak significantly at $p_2^0 = \pm |\vec{p_2}|$, we could apply the approximation to regard them as massless bosons. The corresponding Feynman diagrams are the same as the second panels in Fig.~\ref{NWl},~\ref{NZv} with the only difference that the Goldstone boson's components are no longer bounded with the longitudinal polarizations of the $W$ and $Z$ bosons.

The dispersion relation of a ``massless'' Goldstone boson is simple,
\begin{eqnarray}
F_{G} (p_2) = (p_2^0)^2 - \vec{p}_2^2 = 0. \label{GoldstoneDispersion}
\end{eqnarray}
Other equations are the same as the previous subsections. After solving  (\ref{LDispersion}, \ref{EConserve}, \ref{pConserve}) with (\ref{GoldstoneDispersion}), we then write down the final result of the squared amplitude
\begin{eqnarray}
A_{G^{\pm}} = \sum_{r,s=1,2} \mathcal{M}_{W}^4 \mathcal{M}_{W}^{* 4} f_F(\frac{p_1^0}{T} ) f_B(\frac{p_2^0}{T}) Z_l(p_1) Z_{G^{\pm}}(p_2)
\end{eqnarray}
for the charged Goldstone channel, where $Z_{G^{\pm}}(p_2)$ is calculated and defined by
\begin{eqnarray}
Z_{G^{\pm}}(p_2) = \frac{2 | \vec{p}_2 |}{\pi} \int_{0}^{|\vec{k}|+\delta} \text{Im}[ i \frac{p_2^2-\Pi_L^W(p_2)+i \epsilon}{p_2^2-[m_W(T)]^2-\Pi_L^W(p_2)+i \epsilon} \frac{i}{p_2^2+i \epsilon}] dk^0,
\end{eqnarray}
and the final result
\begin{eqnarray}
A_{G^{0}} = \sum_{r,s=1,2} \mathcal{M}_{Z\text{/}\gamma}^4 \mathcal{M}_{Z\text{/}\gamma}^{* 4} f_F(\frac{p_1^0}{T} ) f_B(\frac{p_2^0}{T}) Z_l(p_1) Z_{G^0}(p_2)
\end{eqnarray}
for the neutral Goldstone channel, where
\begin{eqnarray}
& & Z_{G^0}(p_2) = \nonumber \\
& & \frac{2 |\vec{p}_2|}{\pi} \int_{0}^{|\vec{k}|+\delta} \text{Im}[ i \frac{(p_2^2-\Pi_L^{11} + i \epsilon)(p_2^2 - \Pi_L^{22} + i \epsilon) - (\Pi_L^{12})^2}{(p_2^2 - [m_Z (T)]^2-\Pi_L^{11} + i \epsilon)(p_2^2 - \Pi_L^{22} + i \epsilon) - (\Pi_L^{12})^2} \frac{i}{p_2^2 + i \epsilon}.
\end{eqnarray}
Here $m_Z(T) = \frac{\sqrt{g_1^2+g_2^2}}{2} v(T)$, and
\begin{eqnarray}
\Pi_L^{11} &=& \Pi_{L}^{B} \sin^2 \theta_W + \Pi_{L}^{W} \cos^2 \theta_W, \nonumber \\
\Pi_L^{22} &=& \Pi_{L}^{B} \cos^2 \theta_W + \Pi_{L}^{W} \sin^2 \theta_W, \nonumber \\
\Pi_L^{12} &=& \Pi_{L}^{W} \cos \theta_W \sin \theta_W - \Pi_{L}^{B} \cos \theta_W \sin \theta_W.
\end{eqnarray}

\subsection{Higgs channels}

\begin{figure}
\includegraphics[width=2in]{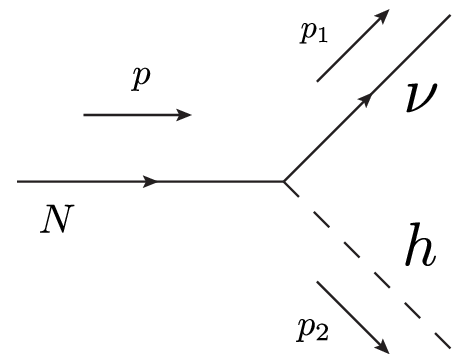}
\caption{$N \rightarrow h \nu$ $1\leftrightarrow2$ channel.} \label{Nvh}
\end{figure}
The Higgs channel is quite straightforward, since the Higgs boson only receives a trivial mass correction from the thermal environment. Below the $T_c$, $m_h(T) \propto v(T)$, so
\begin{eqnarray}
m_h(T) = m_{h0} \sqrt{1-\frac{T^2}{T_c^2}},
\end{eqnarray}
and above the $T_c$, $m_h(T)$ becomes
\begin{eqnarray}
m_h^2(T) = (g_1^2 + 3 g_2^2 + 4 y_t^2 + 8 \lambda) \frac{T^2-T_c^2}{16},
\end{eqnarray}
where $m_{h0} = 125 \text{ GeV}$. Therefore the dispersion relation of a Higgs boson is simply
\begin{eqnarray}
F_{H}(p_2) =p_2^2 - m_h(T)^2 = 0.	\label{HiggsDispersion}
\end{eqnarray}
Again solving (\ref{LDispersion}, \ref{EConserve}, \ref{pConserve}) with (\ref{HiggsDispersion}) for the valid $p_1$ and $p_2$, we then write down the amplitude,
\begin{eqnarray}
i \mathcal{M}_{h} = i y_N \bar{u}^s (\tilde{p}_1) P_R u^r(p).
\end{eqnarray}
The total result of the squared amplitude is
\begin{eqnarray}
A_h = \sum_{r,s=1,2} \mathcal{M}_h \mathcal{M}^*_h f_F(\frac{p_1^0}{T}) f_B(\frac{p_2^0}{T}) Z_l(p_1).
\end{eqnarray}

\section{Phase Space and Thermal Average Integration}	\label{Integration}

In the thermal background, the Lorentz invariance is broken so that we could not directly ``boost'' the center of momentum reference frame to calculate the $1\leftrightarrow2$ processes of a sterile neutrino at rest. We could only rely on the definition of a width at an arbitrary reference frame
\begin{eqnarray}
\Gamma_X &=& \frac{1}{2 p^0} \int \frac{d^3 \vec{p}_1 d^3 \vec{p}_2}{(2 \pi)^6} \frac{A_X}{(2 p_1^0)(2 p_2^0)} (2 \pi)^4 \delta^4(p-p_1-p_2) \nonumber \\
&=& \frac{1}{2 p^0} \int \frac{d^3 \vec{p}_1}{(2 \pi)^6} \frac{A_X}{(2 p_1^0)(2 p_2^0)} (2 \pi)^4 \delta(p^0-p_1^0-p_2^0) \nonumber \\
&=& \frac{1}{2 p^0} \int \frac{2 \pi \sin \theta_p \vec{p}_1^2 d |\vec{p}_1| d \theta_p}{(2 \pi)^6} \frac{A_X}{(2 p_1^0)(2 p_2^0)} (2 \pi)^4 \delta(p^0-p_1^0-p_2^0)
\end{eqnarray}
where $X=[W,(T,L\text{out})],[Z\text{/}\gamma,(T,L\text{out})],G^{\pm},G^0,h$. Note that in the thermal plasma rest frame, there is still the symmetry of the system rotating along the $\vec{p}$ axis, thus eliminating the azimuthal angle $\phi$ to be a $2 \pi$ factor. To integrate out the $\delta$ function, we calculate
\begin{eqnarray}
\frac{\partial p_1^0}{\partial |\vec{p}_1|} + \frac{\partial p_2^0}{\partial |\vec{p}_1|} = \frac{\partial p_1^0}{\partial |\vec{p}_1|} + \frac{\partial p_2^0}{\partial |\vec{p}_2|}  \frac{\partial |\vec{p}_2|}{\partial |\vec{p}_1|}. \label{PhaseSpace1}
\end{eqnarray}
$\frac{\partial |\vec{p}_2|}{\partial |\vec{p}_1|}$ is extracted from the momentum conservation law (\ref{pConserve}), and the result is
\begin{eqnarray}
\frac{\partial |\vec{p}_2|}{\partial |\vec{p}_1|} = \frac{|\vec{p}_1| - |\vec{p}| \cos \theta_p}{|\vec{p}_2|}.
\end{eqnarray}
$\frac{\partial p_1^0}{\partial |\vec{p}_1|} $ and $\frac{\partial p_2^0}{\partial |\vec{p}_2|}$ can be extracted from the corresponding dispersion relations (\ref{WDispersion}, \ref{DispersionZGamma}, \ref{GoldstoneDispersion}, \ref{HiggsDispersion}). Generally, if the dispersion relation of a momentum $p_Y$ is written to be $F_X (p_Y)=F_X (p_Y^0, |\vec{p}_Y|)=0$, where $X=l, [W, (T,L\text{out})],[Z\text{/}\gamma, (T,L\text{out})],G,H$, then
\begin{eqnarray}
\frac{\partial p_Y^0}{\partial |\vec{p}_Y|}  = -\frac{\frac{\partial F_X (p_Y)}{\partial |\vec{p}_Y|}}{\frac{\partial F_X (p_Y)}{\partial p_Y^0}}. \label{DifferentialP}
\end{eqnarray}
Therefore, (\ref{PhaseSpace1}) can be reduced to
\begin{eqnarray}
\Gamma_X = \frac{1}{2 p^0} \int \frac{2 \pi \sin \theta_p \vec{p}_1^2 d \theta_p}{(2 \pi)^6} \frac{A_X}{(2 p_1^0)(2 p_2^0)} \frac{(2 \pi)^4}{\left| \frac{\frac{\partial F_l (p_1)}{\partial |\vec{p}_1|}}{\frac{\partial F_l (p_1)}{\partial p_1^0}} + \frac{\frac{\partial F_X (p_2)}{\partial |\vec{p}_2|}}{\frac{\partial F_X (p_2)}{\partial p_2^0}} \frac{|\vec{p}_1| - |\vec{p}| \cos \theta_p}{|\vec{p}_2|} \right| }. \label{PhaseSpace2}
\end{eqnarray}
The thermal average integration is then simple,
\begin{eqnarray}
\gamma_X = \int \frac{d^3 \vec{p}}{(2 \pi)^3} e^{-\frac{p^0}{T}} \Gamma_X. \label{ThermalAverage}
\end{eqnarray}
This $\gamma_X$ will enter the Boltzmann equation.

Straightforwardly applying (\ref{PhaseSpace2}-\ref{ThermalAverage}) takes a problem. For each $\theta_p$, sometimes there are multiple solutions for the $p_1^0$, $\vec{p}_1$, $p_2^0$, $\vec{p}_2$ values. One reason is that when a particle decays to every direction in its center of momentum frame, and while boosted to the plasma reference frame, one angle can pick up multiple different momentums. To cure this problem, one can adjust the integration order to calculate in the (inverse-)decayed particle's rest frame.

For example, for sterile neutrino's decay process, we rely on the $N$-rest frame by boosting the $p_1$, $p_2$ into $p_{1N}$, $p_{2N}$. We then use $p_{1,2N}$ as the input parameters to solve the various dispersion relations. We then need to calculate the Jacobian and delta function's factors in the new $p_{1N}$, $p_{2N}$ parameters. Take the x-axis along the $\vec{p}$ direction, and without loss of generality, let $\vec{p}_1$ be located in the x-y plain, and we have
\begin{eqnarray}
p_{1N}^0 &=& \gamma (p_{1}^0 - \beta |\vec{p}_{1}| \cos \theta_p), \label{LorentzTrans1}\\
|\vec{p}_{1N}| \cos \theta_N &=& \gamma (|\vec{p}_{1}| \cos \theta_p - \beta p_{1}^0), \label{LorentzTrans2}\\
|\vec{p}_{1N}| \sin \theta_N &=& |\vec{p}_{1}| \sin \theta_p, \label{LorentzTrans3}
\end{eqnarray}
where $\beta = \frac{|\vec{p}|}{p^0}$, $\gamma=\frac{1}{\sqrt{1-\beta^2}}$. A tedious calculation finally shows that
\begin{eqnarray}
\frac{d \theta_N}{d \theta_p} = \frac{\partial \theta_N}{\partial \theta_p} + \frac{\partial \theta_N}{\partial p_1^0} \frac{\partial p_1^0}{\partial \theta_p} + \frac{\partial \theta_N}{\partial |\vec{p}_1|} \frac{\partial |\vec{p}_1|}{\partial \theta_p}, \label{dThetaNdThetaP}
\end{eqnarray}
where
\begin{eqnarray}
\frac{\partial \theta_N}{\partial \theta_p} &=& \frac{\vec{p}_1^2 \gamma(-|\vec{p}_1| + p_1^0 \beta \cos \theta_p) \sin \theta_p}{\left[ \gamma^2 (p_1^0 \beta - |\vec{p}_1| \cos \theta_p)^2 + \vec{p}_1^2 \sin^2 \theta_p \right]^{\frac{3}{2}}} \frac{1}{(-\sin \theta_N)} , \\
\frac{\partial \theta_N}{\partial p_1^0}  &=& \frac{-\vec{p}_1^2 \beta \gamma \sin^2 \theta_p}{\left[ \gamma^2 (p_1^0 \beta - |\vec{p}_1| \cos \theta_p)^2 + \vec{p}_1^2 \sin^2 \theta_p \right]^{\frac{3}{2}}} \frac{1}{(-\sin \theta_N)}, \\
\frac{\partial \theta_N}{\partial |\vec{p}_1|}  &=& \frac{ p_1^0 |\vec{p}_1| \beta \gamma \sin^2 \theta_p}{\left[ \gamma^2 (p_1^0 \beta - |\vec{p}_1| \cos \theta_p)^2 + \vec{p}_1^2 \sin^2 \theta_p \right]^{\frac{3}{2}}} \frac{1}{(-\sin \theta_N)}, \\
\frac{\partial p_1^0}{\partial \theta_p} &=& \frac{|\vec{p}| |\vec{p}_1| \sin \theta_p}{-\frac{\partial |\vec{p}_1|}{\partial p_1^0} |\vec{p}_1| - \frac{\partial |\vec{p}_2|}{\partial p_2^0} |\vec{p}_2| + \frac{\partial |\vec{p}_1|}{\partial p_1^0 } |\vec{p}| \cos \theta_p }, \\
\frac{\partial \vec{p}_1}{\partial \theta_p} &=& \frac{|\vec{p}| |\vec{p}_1| \sin \theta_p}{ -\vec{p}_1 - \frac{\partial \vec{p}_2}{\partial p_2^0} \frac{\partial p_1^0}{\partial |\vec{p}_1|} |\vec{p}_2| + |\vec{p}| \cos \theta_p}, \label{pp1Overpthetap}
\end{eqnarray}
and $\frac{\partial |\vec{p}_i|}{\partial p_i^0}$ has been already calculated in (\ref{DifferentialP}). Then we can replace the $d \theta_p$ with $d \theta_N \frac{d \theta_p}{d \theta_N}$ in (\ref{PhaseSpace2}) to calculate this integral.

Inverse-decay processes are similar. For example, if we calculate the W-boson's inverse decay process $N l^+ \rightarrow W^+$, we need to adjust the integration order of  (\ref{PhaseSpace2}-\ref{ThermalAverage}) to integrate out the $d^3 \vec{p}$ and $d^3 \vec{p}_1$ phase space at first and finally calculate the $d^3 \vec{p}_2$ integration. Boost to the $W^{+}$'s rest frame to transfer to the $\vec{p}_W$, $\vec{p}_{1W}$ integration by replacing the corresponding indices in the Eqs.~(\ref{LorentzTrans1}-\ref{pp1Overpthetap}) to calculate the similar Jacobian and delta function's factors. With this method, all the $1 \leftrightarrow 2 $ channels can be computed.

Let us summarize the numerical algorithm processes. To calculate one channel, e.g., $N \leftrightarrow Wl$, one needs to follow these steps:
\begin{itemize}
\item Fixing the $p^0$, $\vec{p}$, and $\theta_N$, we are going to solve the $p_{1N}^0$, $p_{2N}^0$, $\vec{p}_{1N}$, $\vec{p}_{2N}$. The equations to be solved are (\ref{WDispersion}, \ref{LDispersion}, \ref{EConserve}, \ref{pConserve}). They are defined with the parameters $p_1^0$, $p_2^0$, $\vec{p}_1$, $\vec{p}_2$ and $\theta_p$, and these two sets of parameters are mediated by (\ref{LorentzTrans1}-\ref{LorentzTrans3}).
\item With the acquired numerical solution of $p_1$, $p_2$ and $\theta_p$, calculate the total squared amplitude through (\ref{WTotalSquaredAmplitude}).
\item Changing $\theta_N$, utilize (\ref{PhaseSpace2}, \ref{dThetaNdThetaP}) to compute the $\Gamma_{W,T\text{/}L}$.
\item Change $p$ to calculate (\ref{ThermalAverage}).
\end{itemize}
To calculate e.g., the $Nl \leftrightarrow W$ channel, we need to integrate out the $\vec{p}$ and $\vec{p}_1$ at first. Thus, exchange the $p$ and $p_2$ in the above items, and also change the subscript $N$ into $W$. Therefore, we are also able to calculate the inverse-decay rate of a $W$ boson below its threshold.

\section{Numerical Results}	\label{NumericalResults}

\begin{table}
\begin{tabular}{|c|c||c|c|}
\hline
Alias & Meaning & Alias & Meaning\\
\hline
WTD & $N \leftrightarrow W_{T}^+ l^-$ & Z$\gamma$LID & $N\bar{v} \leftrightarrow Z_L\text{/}\gamma_L$\\
\hline
WTID & $N l^+ \leftrightarrow  W_{T}^+$ & $G^{\pm}$D & $N \leftrightarrow G^+ l^-$\\
\hline
WLD & $N \leftrightarrow W_{L}^+ l^-$ & $G^{\pm}$ID & $N G^- \leftrightarrow l^-$, $N l^+ \leftrightarrow G^+$ \\
\hline
WLID & $N l^+ \leftrightarrow  W_{L}^+$ & $G^0$D & $N G^0 \leftrightarrow \nu$, $N \bar{\nu} \leftrightarrow G^0$ \\
\hline
Z$\gamma$TD & $N \leftrightarrow Z_T\text{/}\gamma_T \nu$ & $G^0$ID & $N G^0 \leftrightarrow \nu$, $N \bar{\nu} \leftrightarrow G^0$ \\
\hline
Z$\gamma$TID & $N \bar{\nu} \leftrightarrow Z_T\text{/}\gamma_T$ & HD & $N \leftrightarrow h \nu$ \\
\hline
Z$\gamma$LD & $N \leftrightarrow Z_L\text{/}\gamma_L \nu$ & HID & $N h \leftrightarrow \nu$, $N \bar{\nu} \leftrightarrow h$ \\
\hline
\end{tabular}
\caption{Channels to be plotted and their meanings.}\label{ChannelsTab}
\end{table}

We have scanned the $m_N \in [50,200]$ GeV range by an interval of $1$ GeV. For the leptonic sector, both ``particle'' and ``hole'' channels had been included. For the bosonic sector, all the transverse, longitudinal vector bosons, and the Goldstone, Higgs channels had been considered. We have enumerated all the $1 \leftrightarrow 2$ possibilities, however, it is unnecessary to plot all of them. We sum over the results into 14 channels, and show the meaning of them in Tab.~\ref{ChannelsTab}. Notice that the channel $N (W\text{/}Z) \leftrightarrow l^-\text{/}\nu$ is kinematically forbidden in our interested parameter space, so that they are neglected. Compared with the production rate $\gamma_X$, it is more convenient to use the averaged decay width
\begin{eqnarray}
\bar{\Gamma}_X = \frac{g_N \gamma_X}{n_N^{\text{eq}}} = \frac{\int \frac{d^3 \vec{p}}{(2 \pi)^3} e^{-\frac{p^0}{T}} \Gamma_X}{2 \frac{m_N^2 T}{2 \pi^2} K_2(\frac{m_N}{T})},
\end{eqnarray}
where $g_N$ is the degree of freedom of the sterile neutrino, and is cancelled by the same factor in $n_{N}^{\text{eq}}$. The comparison of this parameter with the Hubble constant $H \simeq 1.66 \sqrt{g_*} \frac{T^2}{M_{\text{pl}}}$ can help us judge whether the sterile neutrino starts to deviate from the thermal equilibrium conveniently.

\begin{figure}
\includegraphics[width=0.45\textwidth]{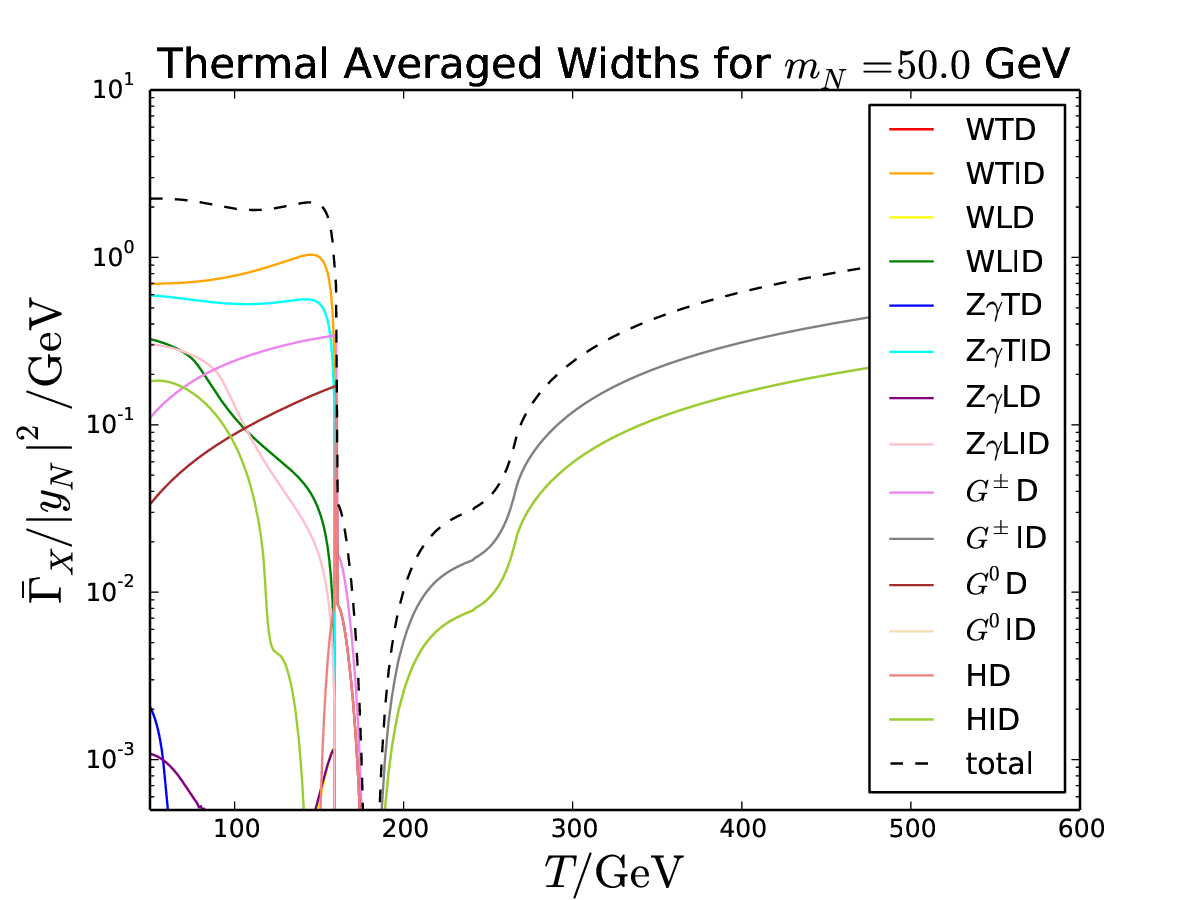}
\includegraphics[width=0.45\textwidth]{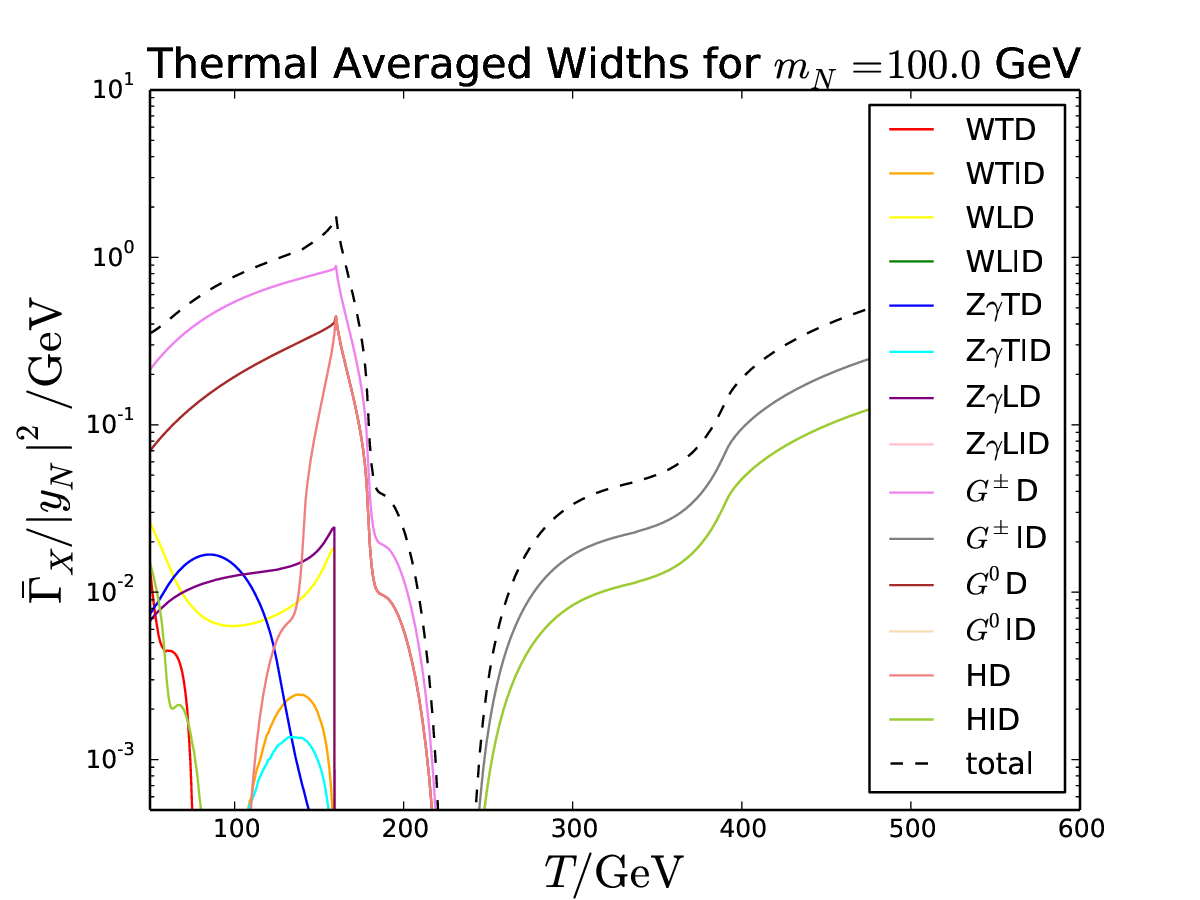}
\includegraphics[width=0.45\textwidth]{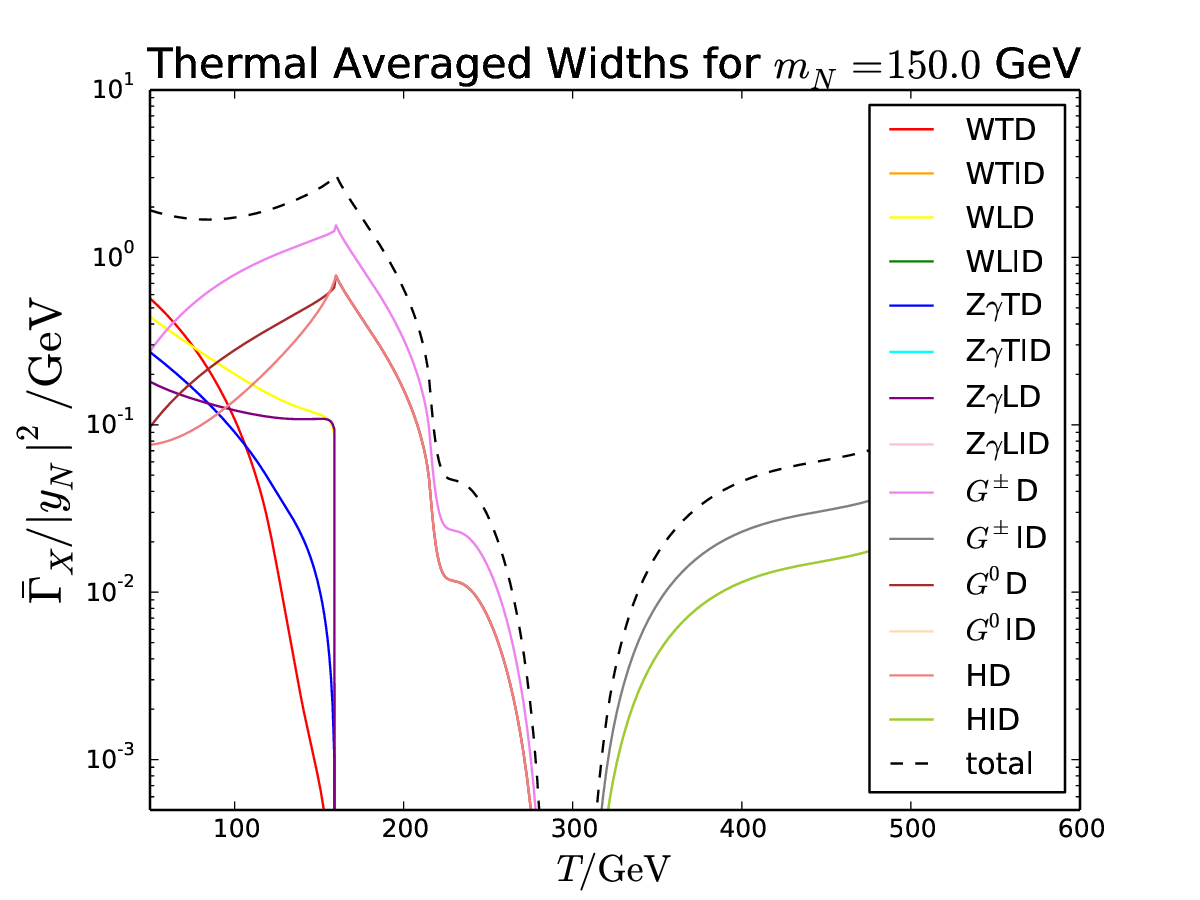}
\includegraphics[width=0.45\textwidth]{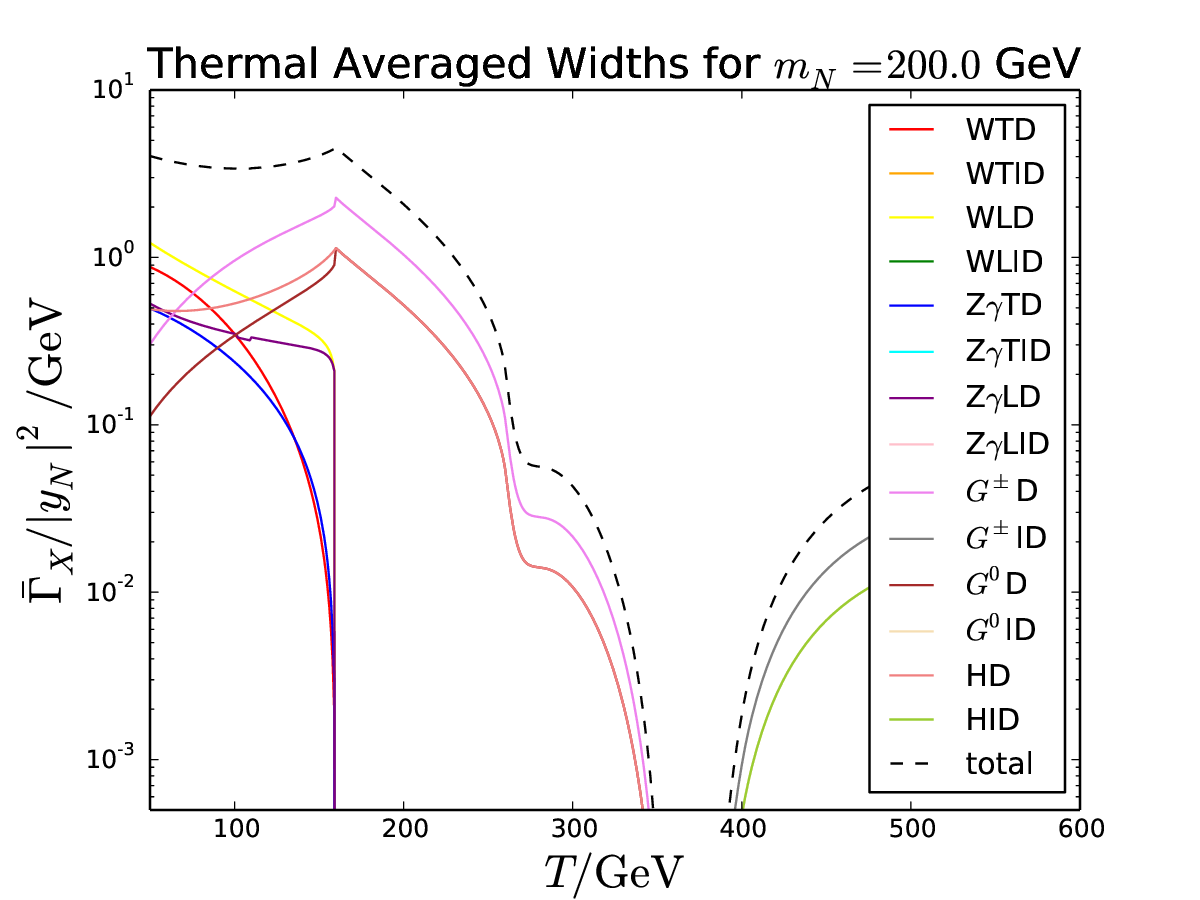}
\caption{Thermal averaged widths plot normalized by $\frac{1}{|y_N|^2}$ for $m_N=50$, $100$, $150$, $200$ GeV masses. The meanings in the legends are illustrated in Tab.~\ref{ChannelsTab}.}\label{GammaXPlots}
\end{figure}

\begin{figure}
\includegraphics[width=0.55\textwidth]{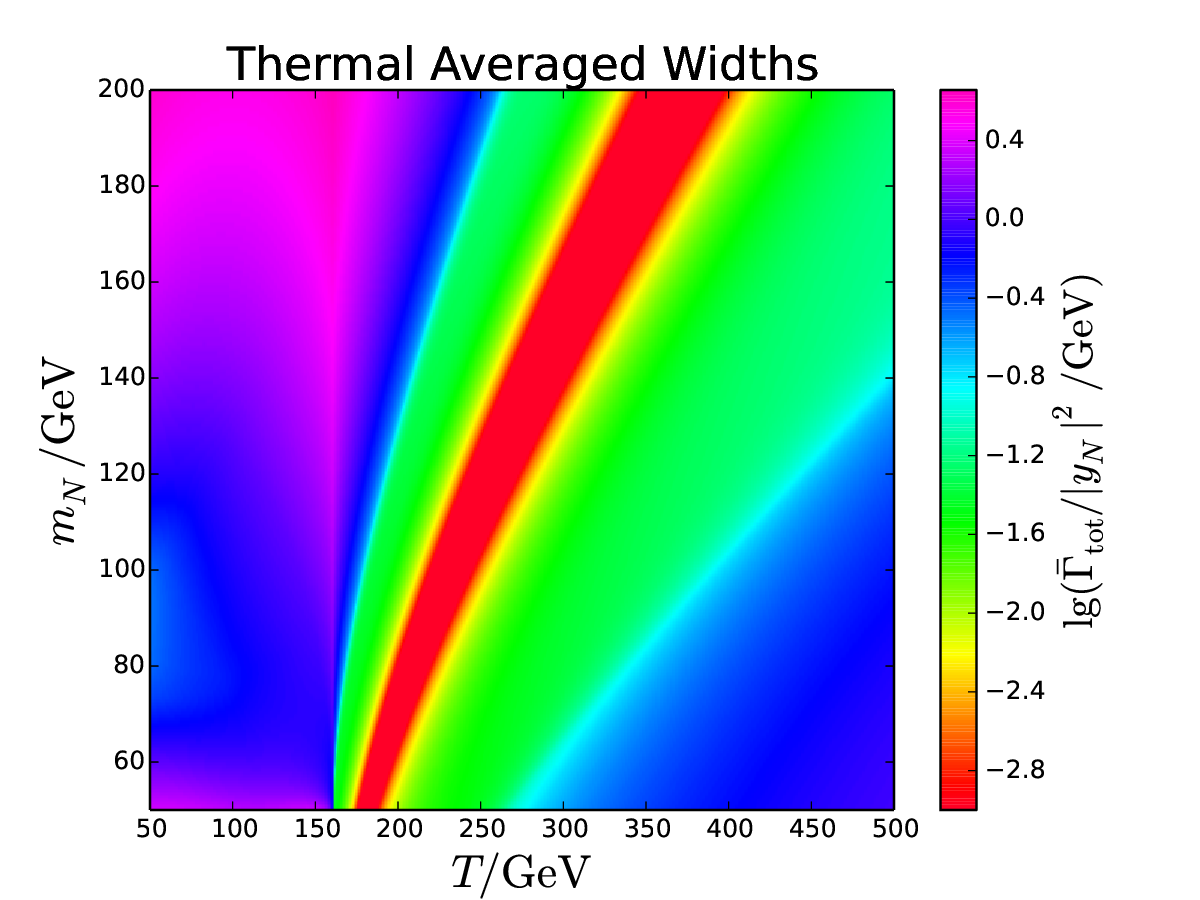}
\caption{$\bar{\Gamma}_{\text{tot}} = \sum\limits_X \bar{\Gamma}_{X}$ normalized by $\frac{1}{|y_N|^2}$ depending on the temperature $T$ and the sterile neutrino mass $m_N$. To keep the image contrast in the other area where the $1 \leftrightarrow 2$ processes are not suppressed kinematically, we just keep $\frac{\bar{\Gamma}_{\text{tot}}}{|y_N|^2} \geq 1.0 \times 10^{-3} \text{ GeV}$ in this image. Therefore, most of the red parts in this image are actually much smaller than those plotted here.} \label{TotalGamma}
\end{figure}

In Fig.~\ref{GammaXPlots}, we have selected the $m_N=50$, $100$, $150$, $200$ GeV to plot their thermal averaged widths normalized by $\frac{1}{|y_N|^2}$ depending on the temperature $T$. Just below the critical temperature $100 \text{ GeV} \lesssim T<T_c$, the longitudinal $W$/$Z$ and the Goldstones play crucial roles. These two kinds of channels are complementary, and can be compared with the corresponding part of Fig.~1 in Ref.~\cite{ThomasPRL}, in which large areas had been kinematically forbidden within the $100 \text{ GeV} \lesssim T<T_c$, $50 \text{ GeV} \lesssim m_N \lesssim 100 \text{ GeV}$ ranges. Our calculations do not give such a remarkable suppression. To show this clearly, we also plot a total thermal averaged width $\bar{\Gamma}_{\text{tot}} = \sum\limits_X \bar{\Gamma}_{X}$ in Fig.~\ref{TotalGamma}. There we can see a similar suppression of the total thermal averaged width when $T>T_c$ compared with the Fig.~1 in Ref.~\cite{ThomasPRL}, while when $T<T_c$, only a slight and obscure suppression appears in roughly the same area.

\begin{figure}
\includegraphics[width=4in]{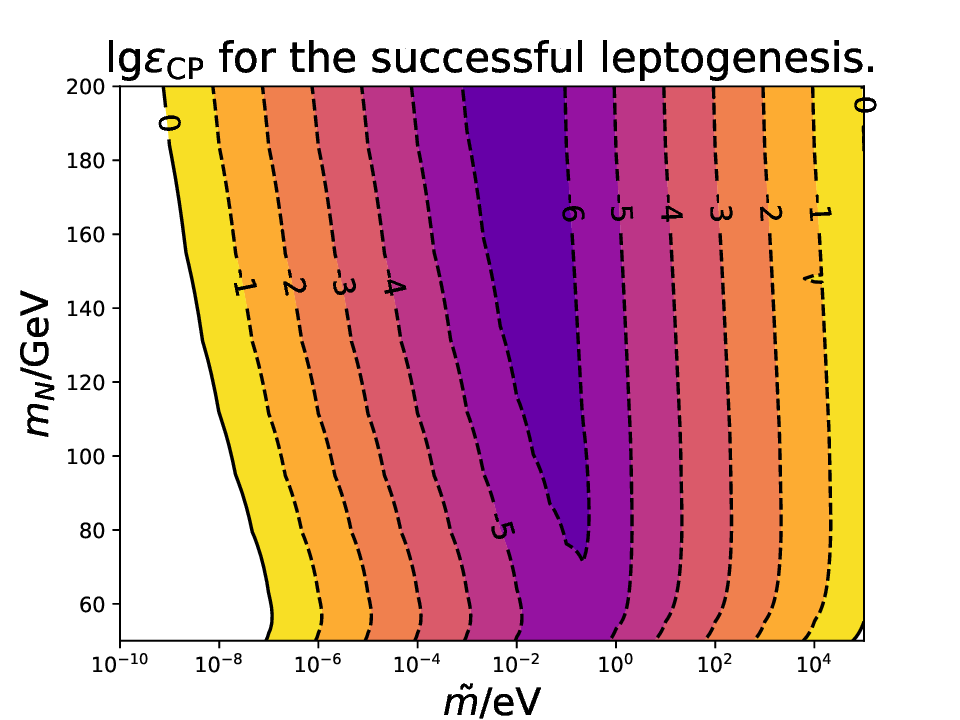}
\caption{$\lg \epsilon_{\text{CP}}$ needed to obtain the successful leptogenesis. The sterile neutrinos are initially in thermal equilibrium with the plasma.} \label{Leptogenesis}
\end{figure}

In the rest of this section we show a preliminary calculation of the leptogenesis with all the results above. Above the sphaleron decoupling temperature, i.e., when $T > T_{\text{sph}} = 131.7$ GeV\cite{Tsph}, the $B+L$ number does not conserve, so the lepton number asymmetry generated from the sterile neutrino $1 \leftrightarrow 2$ processes is ported to the baryon number asymmetry through the sphaleron effects. To explain the observed ratio of baryon asymmetry normalized by the photon number density $|\eta_{B 0}| = \frac{|n_{B}-n_{\bar{B}}|}{n_{\gamma}} \approx 6 \times 10^{-10}$ in our current universe, $|\eta^L| = \frac{|n_{L} - n_{\bar{L}}|}{n_{\gamma}} $ is calculated then to be $2.47 \times 10^{-8}$\cite{Resonant5} at $T = T_{\text{sph}} = 131.7$ GeV. Including the $2 \leftrightarrow 2$ wash-out terms, the Boltzmann equations are given by
\begin{eqnarray}
\frac{n_{\gamma} H_N}{z} \frac{d \eta_N}{d z} &=& \left( 1-\frac{\eta_N}{\eta_N^{\text{eq}}} \right) [\gamma_D + 2(\gamma_{Hs} + \gamma_{As}) + 4 (\gamma_{Ht} + \gamma_{At}) ], \nonumber \\
\frac{n_{\gamma} H_N}{z} \frac{d \eta_L}{d z} &=& \gamma_D \left[ \left( \frac{\eta_N}{\eta_N^{\text{eq}}} -1\right) \epsilon_{\text{CP}}(z) - \frac{2}{3} \eta_L \right] - \frac{4}{3} \eta_L \left[ 2 (\gamma_{Ht} + \gamma_{At}) + \frac{\eta^N}{\eta^N_{\text{eq}}} (\gamma_{Hs}+\gamma_{As}) \right],
\end{eqnarray}
where $\eta_N = \frac{n_{N}}{n_{\gamma}}$, $z=\frac{n_N}{T}$, and $\gamma_D = \sum\limits_X \gamma_X$ is the summation over all the $1 \leftrightarrow 2$ channels defined in (\ref{ThermalAverage}). We shall neglect the $2 \leftrightarrow 2$ contributions $\gamma_{Hs, Ht, As, At}$ in this paper, since we only calculate the situation that the sterile neutrino is initially in thermal equilibrium with the plasma when $T \gg m_N$. When $T \sim m_N$ or $T \lesssim m_N$ that the deviation from the thermal equilibrium becomes significant, the $2 \leftrightarrow 2$ processes are usually suppressed by an additional $n_{A,H,...}^{eq}$ factor compared with $\gamma_D$. The CP-source parameter $\epsilon_{\text{CP}}(z)$ originate from the one-loop interference with the tree-level amplitudes\cite{Resonant1, ResonantEpsilon}, and should depend on $z$. The identification of this parameter is beyond the scope of this paper. We only follow the section II of Ref.~\cite{ThomasPRL} to regard $\epsilon_{\text{CP}}$ as a constant parameter to present our results of the successful leptogenesis in Fig.~\ref{Leptogenesis}. Studies on At some proposed future leptonic colliders, with the aid of the secondary vertex detection, the sensitivity to $y_N$ at ILC\cite{ILC1, ILC2, ILC3, ILC4, ILC5}, CEPC\cite{CEPC1, CEPC2} and FCC-ee\cite{FCC-ee} can be significantly improved. Refs.~\cite{Collider1, Collider2, Collider3, Collider4, Collider5} have discussed the corresponding searches at these colliders, and Ref.~\cite{MondalLHeC, Das2} have also discussed the proposals at the LHeC\cite{LHeC1, LHeC2}, Ref.~\cite{Das1} have discussed the similar parameter space at the LHC and beyond. Their results can roughly verify the parameter space within $50 \text{ GeV} < m_N < 90 \text{ GeV}$ and $\tilde{m} \gtrsim 1 \text{ eV}$. Our contours are significantly different with the Fig.~3 in Ref.~\cite{ThomasPRL},  especially for the $1\text{ eV} \lesssim \tilde{m} \lesssim 10^5 \text{ eV}$ and $40 \text{ GeV} \lesssim m_N \lesssim 110 \text{ GeV}$ area there, where quite a large void appeared due to the absence of the $\gamma_D$ kinematically forbidden below $T_c$ in their Fig.~1. In our paper, such an area is filled up with the $N \leftrightarrow G^{+,0} (l^{-}\text{/}\nu)$, $N \leftrightarrow W^+_{T,L} l^{-}$ or $N \leftrightarrow Z_{T,L} \nu$ channels, so that no significant distortions of the contours appear.

\section{Summary}	\label{Summery}

We have calculated the $1 \leftrightarrow 2$ processes of a sterile neutrino interacting with the gauge/Higgs bosons and leptons in the thermal plasma. We applied the Goldstone-equivalence gauge to evaluate the processes below the critical temperature $T_c \approx 160 \text{ GeV}$, and our method is suitable for the sterile neutrino's mass $m_N \sim T_c$. The results can be utilized in the studies involving the sterile neutrinos, and we have preliminarily calculated the leptogenesis as an example. Compared with Ref.~\cite{ThomasPRL}, the results had been significantly changed due to the different kinematic threshold understandings in this paper. $1 \leftrightarrow 2$ results are usually sufficient to study the processes in the temperature that is roughly of the same magnitude of the sterile neutrino's mass if one assumes an initially thermal equilibrium. Yet the non-perturbative corrections that the leptons and bosons interchange soft particles with the plasma and with each other have not been included. To carry forward our research to a wider temperature scale and to a more precise calculation, we will include all these effects in our further studies. 

\appendix
\section{Aspect from the $R_{\xi}$ gauge}

The advantage of the Goldstone equivalent gauge is the anatomy of the longitudinal polarization and the remained Goldstone degrees of freedom contributions,  which is convenient for one to follow a ``tree-level'' methodology.  The result should be numerically equivalent to the traditional aspect to calculate the imaginary part of the one-loop propagators. In fact,  we showed in Ref.~\cite{VectorBosonMyPaper} that similar ``tree-level'' logic can also be applied in the standard $R_{\xi}$ gauge if only the remained Goldstone degree of freedom is replaced by a ``vector boson'' with the polarization vector $\propto p$, where $p$ is the ``vector boson'''s momentum.  {The equivalence of the results with different gauges is guaranteed by the Ward-Takahashi identity in the broken phase\cite{WTIdentityBrokenPhase}, 
\begin{eqnarray}
p_2^{\mu} \mathcal{M}_{V \mu} = i m_{V}(T) \mathcal{M}^{\text{GS}},	\label{WTI}
\end{eqnarray}
where V=Z\text{/}W, $m_{V}(T)$ is the gauge boson's mass originate from the vev, and $\mathcal{M}^{\text{GS}}$ is the amplitude with the corresponding gauge boson replaced by a Goldstone external leg.  For the W boson, just notice that the relationship between the polarization vectors under two gauges,
\begin{eqnarray}
\epsilon_{L, R_{\xi}}^{W} = \epsilon_{L\text{in}}^{W} + \left( \begin{array}{c}
\frac{p_2^{\mu}}{\sqrt{p_2^2}} \\
-i \frac{m_W(T)}{\sqrt{p_2^2}}
\end{array} \right),
\end{eqnarray}
where $\epsilon_{L, R_{\xi}}^{W}$ is the familiar polarization vector in the $R_{\xi}$ gauge.  One immediately finds out that the contribution from the difference between these two polarization vectors should always vanish according to the Ward-Takahashi identity in the broken phase. 

For the mixing $Z\text{/}\gamma$ case, things are a little bit complicated. Notice that in the (\ref{MZV}), the mixing parameter $-x_1 \sin \theta_W + x_2 \cos \theta_W$ factor is in the vertex term, while (\ref{LongitudinalPolarizationZGamma}), the exactly same thing is attributed to the polarization vector. Remember also for a pure $\gamma$, it does not receive any mass from the vev so its amplitude completely disappears when dotted by the $p_2^{\mu}$.  Factoring out the common $-x_1 \sin \theta_W + x_2 \cos \theta_W$ term, one find that the contribution from the difference between the two polarization vectors
\begin{eqnarray}
\epsilon_{L, R_{\xi}}^{Z\text{/}\gamma} - \epsilon_{L\text{in}}^{Z\text{/}\gamma} = \left( \begin{array}{c}
\frac{p_2^{\mu}}{\sqrt{p_2^2}} \\
-i (-x_1 \sin \theta_W + x_2 \cos \theta_W) \frac{m_Z(T)}{\sqrt{p_2^2}}
\end{array} \right),
\end{eqnarray}
still vanishes in the amplitude, which is also guaranteed by the (\ref{WTI})

The above discussions only involve the longitudinal polarizations of the vector bosons. For the Goldstone channels,  we have pointed out in Ref.~\cite{VectorBosonMyPaper} that these Goldstone external legs can be replaced by a ``vector boson'' with the polarization vector $\frac{p_2^{\mu}}{i m_V}$, } equivalent to picking up the ``quasi-poles'' corresponding to the $\propto p_2^{\mu} p_2^{\nu}$ terms in the $R_{\xi}$ propagator.

One might notice that the Ward-Takahashi identity is not rigorously satisfied {perturbative} if {one only keeps the tree-level part} in (\ref{MWV}, \ref{MWGS}, \ref{MZV}, \ref{MZGS}). {This can be fixed} if we introduce the hard thermal one-loop corrections to the gauge vertices(Page 161 in Ref.~\cite{ThermalFieldBook}),
\begin{eqnarray}
\Gamma_{\mu}(p, p_1) = m_f^2 \int_{\Omega_{\hat{\vec{k}}}} \frac{d \Omega_{\hat{\vec{k}}}}{4 \pi} \frac{\hat{k}_{\mu} \hat{k}\!\!\!/}{ (p \cdot \hat{k})(p_1 \cdot \hat{k}) }, \label{GaugeVertexCorrection}
\end{eqnarray}
where $m_f$ is again given by (\ref{FermionCorrectionMass}) and $\hat{k}=(1, \hat{\vec{k}})$ and $\hat{\vec{k}} \cdot \hat{\vec{k}} = 1$. {The recovery} of the (\ref{WTI}) can be seen by dotting the $p_2 = p - p_1$ into $\Gamma_{\mu}$,
\begin{eqnarray}
(p-p_1) \cdot \Gamma(p, p_1) = m_f^2 \int_{\Omega_{\hat{\vec{k}}}} \frac{d \Omega_{\hat{\vec{k}}}}{4 \pi} \left[ \frac{\hat{k}\!\!\!/}{p_1 \cdot \hat{k} } - \frac{\hat{k}\!\!\!/}{ p \cdot \hat{k} } \right] = \Sigma(p_1)-\Sigma(p), 
\end{eqnarray}
where $\Sigma(p)$ is the hard thermal one-loop correction on a fermionic propagator of the active neutrino or a charged lepton. These two $\Sigma$'s will help cancel the denominators in the $\frac{i}{p\!\!\!/_{(1)}-\Sigma(p_{(1)})}$ propagators on both sides of the gauge vertex, thus {resuming} the Ward-Takahashi identity in the broken phase.

To analytically calculate the (\ref{GaugeVertexCorrection}), we define the dimensionless $K_{\mu \nu}$ by
\begin{eqnarray}
\Gamma_{\mu}(p, p_1) = \frac{m_f^2 K_{\mu \nu} (a, b) \gamma^{\nu}}{p^0 p_1^0}=\frac{m_f^2 K_{\mu \nu} (\alpha, \beta, \theta_{ab}) \gamma^{\nu}}{p^0 p_1^0},
\end{eqnarray}
where
\begin{eqnarray}
a&=&\frac{p}{p^0} = (1, \alpha \frac{\vec{p}}{|\vec{p}|}),	\nonumber \\
b&=&\frac{p_1}{p_1^0}=(1, \beta \frac{\vec{p}_1}{|\vec{p}_1|}),	\nonumber \\
\theta_{ab}&=&\frac{\vec{p}\cdot\vec{p}_1}{|\vec{p}||\vec{p}_1|}. 
\end{eqnarray}
Therefore,
\begin{eqnarray}
K_{\mu \nu}(\alpha, \beta, \theta_{ab}) =  \int_{\Omega_{\hat{\vec{k}}}} \frac{d \Omega_{\hat{\vec{k}}}}{4 \pi} \frac{\hat{k}_{\mu} \hat{k}_{\nu}}{ (a \cdot \hat{k})(b \cdot \hat{k}) }.	\label{KDefinition}
\end{eqnarray}
Obviously $K_{\mu \nu}=K_{\nu \mu}$, and $K_{\mu}^{\mu}=0$. It is then convenient to decompose the $K_{\mu \nu}$ into a combination of the tensor basis,
\begin{eqnarray}
K_{\mu \nu}(\alpha, \beta, \theta_{ab}) &=& A t_{\mu} t_{\nu} + B (t_{\mu} a_{\nu} + t_{\nu} a_{\mu}) + C (t_{\mu} b_{\nu} + t_{\nu} b_{\mu}) + D a_{\mu} a_{\nu} + E b_{\mu} b_{\nu} \nonumber \\
&+& F (a_{\mu} b_{\nu} + a_{\nu} b_{\mu}) + G l_{\mu} l_{\nu}. \label{KDecomposition}
\end{eqnarray}
Here $t = (1,0,0,0)$ is the reference frame vector of the plasma,  and $l=(0,  \frac{\vec{p} \times \vec{p}_1}{|\vec{p}| |\vec{p}_1|})$ which is the unit vector perpendicular to the two input momenta. One might consider extra basis such as $t_{\mu} l_{\nu}$, $a_{\mu} l_{\nu}$, etc.. However, if we rotate to the frame that $l=(0, 1, 0, 0)$,  $a=(1, 0, 0, \alpha)$, $b=(1, 0, \beta \sin \theta_{ab}, \beta \cos \theta_{ab})$, we find the $t_{\mu} l_{\nu}$ etc.~factors all contain such integrals like $\int_{\Omega_{\hat{\vec{k}}}} \frac{d \Omega_{\hat{\vec{k}}}}{4 \pi} \frac{\hat{k}_1 \hat{k}_{0,2,3}}{ (a \cdot \hat{k})(b \cdot \hat{k}) }$, with the integrand which is odd under the transformation $\hat{k}_1 \rightarrow -\hat{k}_1$. Therefore all these terms vanish.

We then contract the $K_{\mu \nu}$ with the $t_{\mu} t_{\nu}$, $t_{\mu} a_{\nu}$, $t_{\mu} b_{\nu}$, $a_{\mu} a_{\nu}$, $b_{\mu} b_{\nu}$, $a_{\mu} b_{\nu}$, $l_{\nu} l_{\nu}$ to determine the $A$-$G$ coefficients. Together with the traceless condition $K_{\mu}^{\mu}=0$, The expressions are
\begin{eqnarray}
K_{tt} = K_{\mu \nu} t^{\mu} t^{\nu} &=& \int_{\Omega_{\hat{\vec{k}}}} \frac{d \Omega_{\hat{\vec{k}}}}{4 \pi} \frac{1}{ (a \cdot \hat{k})(b \cdot \hat{k}) } = A+2B+2C+D+E+2F, \nonumber \\
K_{ta} = K_{\mu \nu} t^{\mu} a^{\nu} &=& \int_{\Omega_{\hat{\vec{k}}}} \frac{d \Omega_{\hat{\vec{k}}}}{4 \pi} \frac{1}{ b \cdot \hat{k} } = A+(2-\alpha^2)B+(2-\alpha \beta \cos \theta_{ab})C \nonumber \\
&+&(1-\alpha^2)D+(1-\alpha \beta \cos \theta_{ab})E + (2-\alpha^2-\alpha \beta \cos \theta_{ab})F, \nonumber \\
K_{tb} = K_{\mu \nu} t^{\mu} b^{\nu} &=& \int_{\Omega_{\hat{\vec{k}}}} \frac{d \Omega_{\hat{\vec{k}}}}{4 \pi} \frac{1}{ a \cdot \hat{k} } = A+(2-\alpha \beta \cos \theta_{ab})B + (2-\beta^2) C \nonumber \\
&+& (1-\alpha \beta \cos \theta_{ab}) D + (1-\beta^2) E + (2-\beta^2-\alpha \beta \cos \theta_{ab}) F, \nonumber \\
K_{aa}=K_{\mu \nu} a^{\mu} a^{\nu} &=& \int_{\Omega_{\hat{\vec{k}}}} \frac{d \Omega_{\hat{\vec{k}}}}{4 \pi} \frac{a \cdot \hat{k} }{ b \cdot \hat{k} } = A+2(1-\alpha^2)B+2(1-\alpha \beta \cos \theta_{ab})C \nonumber \\
&+& (1-\alpha^2)^2 D + (1-\alpha \beta \cos \theta_{ab})^2 E + 2(1-\alpha^2)(1-\alpha \beta \cos \theta_{ab}) F, \nonumber \\
K_{bb}=K_{\mu \nu} b^{\mu} b^{\nu} &=& \int_{\Omega_{\hat{\vec{k}}}} \frac{d \Omega_{\hat{\vec{k}}}}{4 \pi} \frac{b \cdot \hat{k} }{ a \cdot \hat{k} } = A+2(1-\alpha \beta \cos \theta_{ab})B+2(1-\beta^2)C \nonumber \\
&+& (1-\alpha \beta \cos \theta_{ab})^2 D + (1-\beta^2)^2 E + 2(1-\beta^2)(1-\alpha \beta \cos \theta_{ab}) F, \nonumber \\
K_{ab}=K_{\mu \nu} a^{\mu} b^{\nu} &=& \int_{\Omega_{\hat{\vec{k}}}} d \Omega_{\hat{\vec{k}}} = A+(2-\alpha^2-\alpha \beta \cos \theta_{ab})B + (2-\beta^2-\alpha \beta \cos \theta_{ab})C \nonumber \\
&+& (1-\alpha^2)(1-\alpha \beta \cos \theta_{ab})D + (1-\beta^2)(1-\alpha \beta \cos \theta_{ab})E \nonumber \\
&+& [(1-\alpha^2)(1-\beta^2)+(1-\alpha \beta \cos \theta_{ab})^2]F, \nonumber \\
K_{ll}=K_{\mu \nu} l^{\mu} l^{\nu} &=& \int_{\Omega_{\hat{\vec{k}}}} \frac{d \Omega_{\hat{\vec{k}}}}{4 \pi} \frac{(l \cdot \hat{k})(l \cdot \hat{k}) }{ (a \cdot \hat{k})(b \cdot \hat{k}) } = G.   \nonumber \\
K_{\mu}^{\mu} = A+2 B &+& 2 C+(1-\alpha^2) D+(1-\beta^2) E+(2-2 \alpha \beta \cos \theta_{ab}) F -G=0\label{Decomposition}
\end{eqnarray}

It is convenient to calculate all the integrals in (\ref{Decomposition}) within the $l=(0, 1, 0, 0)$,  $a=(1, 0, 0, a)$, $b=(1, 0, b \sin \theta_{ab}, b \cos \theta_{ab})$ framework. We just list the results below,
\begin{eqnarray}
K_{tt} &=& \frac{\mathrm{artanh} \left[ \frac{2(1-\alpha \beta \cos \theta_{ab}) \sqrt{\alpha^2+\beta^2-\alpha^2 \beta^2+\alpha \beta \cos \theta_{ab}(\alpha \beta \cos \theta_{ab}-2)}}{1+\alpha^2+\beta^2-4 \alpha \beta \cos \theta_{ab}+\alpha^2 \beta^2 \cos 2 \theta_{ab}} \right]}{2 \sqrt{\alpha^2+\beta^2-\alpha^2 \beta^2+\alpha \beta \cos \theta_{ab}(\alpha \beta \cos \theta_{ab}-2)}}, \nonumber \\
K_{ta} &=& \frac{\mathrm{artanh} \beta}{\beta}, ~~K_{tb}=\frac{\mathrm{artanh} \alpha}{\alpha}, \nonumber \\
K_{aa} &=& \frac{(\beta-\alpha \cos \theta_{ab}) \mathrm{artanh} \beta+\alpha \beta \cos \theta_{ab}}{\beta^2},  \nonumber \\
K_{bb} &=& \frac{(\alpha-\beta \cos \theta_{ab}) \mathrm{artanh} \alpha+\alpha \beta \cos \theta_{ab}}{\alpha^2},  \nonumber \\
K_{ab} &=& 1, \nonumber \\
K_{ll} &=& \frac{\mathrm{artanh} \beta (\beta - \alpha \cos \theta_{ab}) + \mathrm{artanh} \alpha (\alpha-\beta \cos \theta_{ab})}{\alpha^2 \beta^2 \sin^2 \theta_{ab} }  \nonumber \\
&+& \mathrm{artanh} \left[ \frac{(\alpha \beta \cos \theta_{ab}-1) \sqrt{4 \alpha^2 + 4 \beta^2 -2 \alpha^2 \beta^2 +2 \alpha \beta(\alpha \beta \cos 2 \theta_{ab}-4 \cos \theta_{ab})}}{1+\alpha^2+\beta^2+\alpha \beta (\alpha \beta \cos 2 \theta_{ab} - 4 \cos \theta_{ab} )} \right]
\nonumber \\
& \times &\frac{\sqrt{4 \alpha^2 + 4 \beta^2 -2 \alpha^2 \beta^2 +2 \alpha \beta(\alpha \beta \cos 2 \theta_{ab}-4 \cos \theta_{ab})}}{4 \alpha^2 \beta^2 \sin^2 \theta_{ab} }. \label{KValues}
\end{eqnarray}
Take (\ref{KValues}) into (\ref{Decomposition}), we acquire eight equations with seven unknown parameters. Solve seven of them to acquire $A$-$G$,  then the $K_{\mu \nu}(\alpha, \beta, \theta_{ab})$ is determined through (\ref{KDecomposition}).

When, however, $\vec{p}$ and $\vec{p}_1$ are nearly parallel to each other, or when one of them are extremely small, the above method suffers from the instability due to the nearly-degeneration of the matrix corresponding to the linear equations in (\ref{Decomposition}). To cure this problem,when $\vec{p}$ and $\vec{p}_1$ are nearly parallel to each other, we estimate the $K_{\mu \nu}$ by taking the $\theta_{ab} \rightarrow 0$ limit, 
\begin{eqnarray}
K_{tt, \theta_{ab} \rightarrow 0}&=&\frac{\mathrm{artanh} \alpha-\mathrm{artanh} \beta}{\alpha - \beta}, \nonumber \\
K_{l l, \theta_{ab} \rightarrow 0}&=&\frac{(\alpha^2-1) \beta^2 \mathrm{artanh}\alpha-\alpha^2 (\beta^2-1) \mathrm{artanh}\beta}{2 \alpha^2 \beta^2 (\alpha-\beta)} - \frac{1}{2 \alpha \beta},
\end{eqnarray}
or taking the $\theta_{ab} \rightarrow \pi$ limit
\begin{eqnarray}
K_{tt, \theta_{ab} \rightarrow \pi}&=&\frac{\mathrm{artanh} \alpha+\mathrm{artanh} \beta}{\alpha + \beta}, \nonumber \\
K_{l l, \theta_{ab} \rightarrow \pi}&=&\frac{(\alpha^2-1) \beta^2 \mathrm{artanh}\alpha+\alpha^2 (\beta^2-1) \mathrm{artanh}\beta}{2 \alpha^2 \beta^2 (\alpha+\beta)} + \frac{1}{2 \alpha \beta}.
\end{eqnarray}
Then $K_{\mu \nu}$ can be expressed as
\begin{eqnarray}
K_{\mu \nu, \vec{p}\parallel \vec{p}_1}(\alpha, \beta, \theta_{a b}) &=& K_{tt} t^{\mu} t^{\nu} + K_{ll} l_{1 \mu} l_{1 \nu} + K_{ll} l_{2 \mu} l_{2 \nu} + K_{\perp} (a-t)_{\mu} (a-t)_{\nu} \nonumber \\
&+& \frac{K_{tt}-K_{ta}}{\alpha^2} [(a-t)_{\mu} t_{\nu} + (a-t)_{\nu} t_{\mu}],
\end{eqnarray}
where $l_1$ and $l_2$ are two unit vectors perpendicular to the $\vec{p}$ without the time component, and also $l_1 \perp l_2$. $K_{\perp}=2 K_{ll}-K_{tt}$ due to the traceless condition.  When, in the other case, and without loss of generality, when $\alpha > \beta$ and $\beta \ll 1$, we can estimate the $K_{\mu \nu}$ by taking the $\beta \rightarrow 0$ limit to acquire
\begin{eqnarray}
K_{tt, \beta \rightarrow 0} &=& \frac{\mathrm{artanh} \alpha}{\alpha}, \nonumber \\
K_{ll, \beta \rightarrow 0} &=& \frac{\alpha+(\alpha^2-1) \mathrm{artanh}(\alpha)}{2 \alpha^3},
\end{eqnarray}
and again
\begin{eqnarray}
K_{\mu \nu, \beta \rightarrow 0}(\alpha, \beta, \theta_{a b}) &=& K_{tt} t^{\mu} t^{\nu} + K_{ll} l_{1 \mu} l_{1 \nu} + K_{ll} l_{2 \mu} l_{2 \nu} + K_{\perp} (a-t)_{\mu} (a-t)_{\nu}  \nonumber \\
&+& \frac{K_{tt}-K_{ta}}{\alpha^2} [(a-t)_{\mu} t_{\nu} + (a-t)_{\nu} t_{\mu}].
\end{eqnarray}

If one wants a gauge invariant result whenever the HTL corrected dispersion relations are considered, (\ref{GaugeVertexCorrection}) should be included.  We can estimate its contributions through a power-counting consideration. Neglecting (\ref{GaugeVertexCorrection}) will introduce a relative error of $ \sim \frac{m_f^2}{m_N^2}$ in the final results. $\frac{m_f^2}{m_N^2} \lesssim 1$ induces $m_N \lesssim 0.15 T$. Since the $W\text{/}Z\text{/}\gamma$ channels open up at $T<T_c$, and $0.15 T_c = 24 \text{ GeV}$. Therefore, our interested range $m_N > 50$ GeV is sufficiently safe if we neglect the vertex thermal correction terms.

The above discussions depends on the assumption that $K_{\mu \nu} \sim 1$. However, the artanh functions in (\ref{Decomposition}) diverge when $\alpha, \beta \rightarrow 1$. This can be realized by observing the denominator of (\ref{KDefinition}), which can be close to zero when $\alpha$, $\beta$ approach 1. Fortunately, this usually happens when a largely boosted ``hole'' is created. The divergence is significantly suppressed by the ``renormaliztion factor'' $Z_{l}(p_1) = \frac{ (p_1^0)^2-\vec{p}_1^2}{2 m_f^2} \propto e^{\frac{-\vec{p}^2}{m_f^2}}$ in (\ref{LeptonRenormalizationFactor}). Therefore, the final integrated rate nearly remains intact, although in this paper we still reckoned in the (\ref{GaugeVertexCorrection}) terms.

In fact, our practical evaluation shows that simpler tree-level vertex method gives not much difference in the final result compared with the data showed in this paper. The Goldstone equivalence gauge also takes another advantage in the tree-level vertex approximation. If we fix on the $R_{\xi}$ gauge, one might introduce a discontinuity of the total effective decay rate over the cross-over temperature $T_c$ up to tree-level. Notice that below the $T_c$, the Goldstone boson fraction's contributions are collected within the $p_2^{\mu} p_2^{\nu}$ terms in the gauge boson components, while when $T>T_c$, all the Goldstone contributions originate from the Yukawa couplings. A continuous transition between these two coupling formalisms requires (\ref{GaugeVertexCorrection}), and neglecting this will introduce a discontinuity. Therefore, we can see that attributing all the ``Goldstone contribution'' of a vector boson to the Goldstone Yukawa couplings, just as what we did in the Goldstone equivalence gauge, will automatically include the key part of the (\ref{GaugeVertexCorrection}) corrections to connect the two parts. Therefore, compared with the $R_{\xi}$ gauge, Goldstone equivalence gauge includes more hard thermal loop corrections on vertices up to a tree-level evaluation.

\begin{acknowledgements}
We thank to Junmou Chen, Pyungwon Ko, Ligong Bian,  Fa-Peng Huang, Chun Liu, Chen Zhang, Ye-Ling Zhou, Mikko Laine, Kechen Wang for helpful discussions and communications. This work is supported in part by the National Natural Science Foundation of China under Grants No. 11805288, No. 11875327 and No.12005312, the Natural Science Foundation of Guangdong Province under Grant No. 2016A030313313, the Fundamental Research Funds for the Central Universities, and the Sun Yat-Sen University Science Foundation. Part of the calculation was performed on TianHe-2, and we thank for the support of National Supercomputing Center in Guangzhou (NSCC-GZ).

\end{acknowledgements}

\appendix

\newpage
\bibliography{SterileNeutrinoDecay}
\end{document}